# Visual Framing of Science Conspiracy Videos: Integrating Machine Learning with Communication Theories to Study the Use of Color and Brightness

Kaiping Chen, Sang Jung Kim, Qiantong Gao, Sebastian Raschka*

**Abstract**

Recent years have witnessed an explosion of science conspiracy videos on the Internet, challenging science epistemology and public understanding of science. Scholars have started to examine the persuasion techniques used in conspiracy messages such as uncertainty and fear yet, little is understood about the visual narratives, especially how visual narratives differ in videos that debunk conspiracies versus those that propagate conspiracies. This paper addresses this gap in understanding visual framing in conspiracy videos through analyzing millions of frames from conspiracy and counter-conspiracy YouTube videos using computational methods. We found that conspiracy videos tended to use lower color variance and brightness, especially in thumbnails and earlier parts of the videos. This paper also demonstrates how researchers can integrate textual and visual features in machine learning models to study conspiracies on social media and discusses the implications of computational modeling for scholars interested in studying visual manipulation in the digital era. The analysis of visual and textual features presented in this paper could be useful for future studies focused on designing systems to identify conspiracy content on the Internet.

**Keywords**. conspiracy, color and brightness, YouTube, computer vision, machine learning, text analysis

---

*Kaiping Chen (http://kaipingchen.com) is an Assistant Professor in Computational Communication in the Life Sciences Communication Department at the University of Wisconsin-Madison and is the corresponding author of this paper (kchen67@wisc.edu). Sang Jung Kim is a PhD student in the School of Journalism and Mass Communication at the University of Wisconsin-Madison. Qiantong Gao is a master's student in data science at Stanford University. Sebastian Raschka (http://pages.stat.wisc.edu/~sraschka/) is an Assistant Professor in the Statistics Department at the University of Wisconsin-Madison. All authors have contributed equally to this project.

Recent years have witnessed an explosion of science conspiracy videos on the Internet, challenging science epistemology and public understanding of science (Ahmed et al., 2020). The U.S. surgeon general issued a public advisory and urged the American society to fight against health misinformation (Murthy, 2021). Despite efforts to debunk conspiracies, they are still believed by many populations. Scholars have started to examine the persuasion techniques used in conspiracy messages such as uncertainty and fear (van Prooijen & Douglas 2017) yet, little is understood about the visual narratives, especially how visual narratives differ in videos that support conspiracy verses those that debunk conspiracy. Having a comparative understanding about the visual features in conspiracies vs debunking videos are crucial as both narratives co-exist and compete in the media ecosystem. Our paper addresses this gap by revealing how conspiracy science videos manipulate colors and brightness to arouse uncertainty and fear. As a result, our work will bring new knowledge to communication theories on framing, uncertainty, and fear appeals, and inform strategies for distinguishing science from misinformation.

To investigate visual framing in conspiracy videos, we integrated communication literature on framing and computational works on visual features used in the entertainment industry. Recent works suggested that visual features of conspiracy theories are similar to those in horror films, sharing the culture of paranoia (Stemmet, 2012). Scholars proposed computable visual features to distinguish horror films from other genres (Rasheed, Sheikh, & Shah, 2005). Among the visual features used in horror films, the use of color and brightness (e.g., lighting and color variances) is found to be closely connected to people's emotion (Valdez & Mehrabian, 1994) and mapped to core concepts in communication, such as appealing to uncertainty and fear (Hollway & Jefferson 2005). Our computational works illuminate a novel approach to study conspiracy theories through exploring the parallel between visual features used in horror films and in conspiracy theories. Through analyzing millions of images from thousands of YouTube videos on COVID-19 conspiracies and climate change conspiracies, we found that conspiracy videos tended to use lower color variances and lighting compared to correction videos, especially in the thumbnails and at the first few seconds of the video. In addition, our paper presents a computational framework for studying images by analyzing to what extent integrating text and visual features will enhance model performance for distinguishing between conspiracy and conspiracy-correction videos, which can yield useful insights for developing classifiers for automatically identifying conspiracy videos in future studies.

## Literature Review

With the rapid circulation of conspiracy theories on social media platforms and their detrimental effects on scientific foundations in society, understanding the nature of conspiracy theories has become a critical topic in communication (Halpern, Valenzuela, Katz, & Miranda, 2019). A conspiracy theory is a type of misinformation that discredits established institutions by building a narrative that these institutions have nefarious intents to threaten society (Scheufele & Krause, 2019). Conspiracy theories challenge established scientific knowledge, such as claiming that 5G technology caused COVID-19 or asserting that scientific institutions have malicious intentions to use the climate change discourse to harm citizens. Such misinformation leads citizens to reject scientific knowledge and threatens informed citizenship, which is crucial for democratic decision-making (Moore, 2018). Conspiracy theories are rapidly circulated on social media platforms and blur the boundaries between authoritative and alternative information sources (Kou, Gui, Chen, & Pine, 2017). Simultaneously, there are corrective efforts from social media users to combat misinformation circulated on social platforms (Vraga, Kim, & Cook, 2019). While the prevalence of conspiracy theories in the social media environment could be damaging, the existence of correction messages combating conspiracy theories by users could contribute to the social media environment's self-purification. However, before comparing the influence of conspiracy theories and corrective messages, it is essential to understand the content characteristics of conspiracy theories and correction messages on social media platforms.

### *Framing Theory and Visual Framing in Communication Field*

Both conspiracy theories and corrective messages on social media platforms attempt to persuade audiences and influence their beliefs. While conspiracy theories aim to convince audiences that established institutions have malicious intentions, correction messages aim to dispel audiences' beliefs in conspiracy theories (Garrett, Nisbet, & Lynch, 2013). We rely on framing theories, a core theory in the communication field to examine persuasive messages, to understand the characteristics of conspiracy theories and correction messages. Message frames presented by media are defined as message attributes that lead audiences to interpret events or issues in a particular way (Chong & Druckman, 2007).

Even though framing studies mostly focused on the textual contexts of messages (Geise, 2017), examining visual frames of messages is gaining importance in the social media era since sharing visuals became an essential part of the social media experience (Filimonov, Russmann, & Svensson, 2016). Visual framing provides audiences analogous image frames that associate directly with certain aspects of reality (Oxman, 2010). Compared to textual frames, audiences can intuitively process visual frames since images resemble what we perceive in our daily life through our eyes (Rodriguez & Dimitrova, 2011). Images are powerful in this aspect, capturing attention easier than texts and transmitting emotion effectively (Brantner, Lobinger, & Wetzstein, 2011). While textual frames utilize words or phrases as framing devices to present causal relationships between events, visual frames use visual modalities or visual distance to make certain aspects of issues salient (Rodriguez & Dimitrova, 2011).

One of the framing devices that enable visual framing is the visual modality (Rodriguez & Dimitrova, 2011). Visual modality refers to "the degree to which certain means of pictorial expression (e.g., color, depth, tonal shades) are used" (Kress & van Leeuwen, 1996, p.256). Visual modality (i.e., visual attributes), such as the degree of color, brightness or saturation, affects individuals' emotions (Valdez & Mehrabian, 1994) or guides individuals to behave in a particular way (Meier, Agostino, Elliot, Maier, & Wilkowski, 2012). Therefore, it is essential to examine the visual modality of persuasive images or videos to understand how they frame specific issues or events. Even though there are studies demonstrating the influence of visual modality on individuals' emotions and behaviors, there is a dearth of studies directly analyzing the visual modalities of images embedded in persuasive messages.

### *Color as Visual Frames and Its Effect on Persuasion*

Visual frames in the form of visual modalities such as degree of color or saturation are potent in persuading audiences (Garber & Hyatt, 2003; Gerend & Sias, 2009; Koh & Cui, 2020). The color of an image delivers symbolic and associative information about the object that appeared in the image (Hine, 1996) and is thought to be the most "resonant and meaningful" visual cue in a perceptual experience (Hilbert, 1987, p.2; Garber & Hyatt, 2003). This attribute of color makes color a captivating visual cue for persuasive communication (Garber & Hyatt, 2003).

At times, viewers' direct sensory experience to the color of an image dominates the cognitive responses of viewers in persuasion (Garber, Hyatt, & Starr, 2000). For example, individuals' initial judgment of a food product is mostly determined by color associated with other sensory cues (e.g., flavor) to accept the product (DuBose, Cardello, & Maller, 1980). In fact, Peng and Jemmott (2018) found that individuals are attracted by food images using arousing colors (e.g., red, orange), and are more likely to share these images online. Caviano and López (2010) demonstrated how color can be intertwined with visual statements and form a visual rhetoric to persuade viewers. Color can function as a "privileged element to 'argue'" in visuals, since colors are often associated with ideas or beliefs about them. For instance, individuals might associate the color of green as natural, red as alerting, and black or darker colors as deadly. As such, color used in images or videos often form its own argument apart from the object portrayed and persuades audiences to perceive an object in a certain way. Examples of this includes anti-war posters using dark colors to depict a war, making viewers to associate war with death.

Koh and Cui (2020) also found that the colorfulness of thumbnails in YouTube videos function as peripheral cues for individuals to click and watch videos. This suggests that color forms its own rhetorical argument that could persuade audiences and make audiences further engage in image or video contents. Moreover, individuals are more likely to be persuaded by the main argument of a visual message when associated with certain colors. Loss-framed messages about HPV vaccination compared to gain-framed messages were only more effective when the color red was primed with loss-framed messages (Gerend & Sias, 2009). This evidence demonstrates how the atmosphere of a message set by the color of an image (or a video) could alter the way audiences think about the message.

Understanding the role of color in persuasion is especially important in misinformation research, since agents spreading misinformation are known to use distinct color tactics than other information sources (Cao et al., 2020; Singh, Ghosh, & Sonagara, 2021). Compared to credible news articles, fake news articles are more likely to use darker images (Singh et al., 2021). Also, misinformation spreaders are known to manipulate visual modalities of content to evoke emotional provocations (Cao et al., 2020). Yet, misinformation studies focusing on color or brightness are still in their infancy and were conducted in computational fields only recently (Cao et al., 2020; Singh et al., 2021). Besides, prominent visual characteristics of conspiracy theories discernible from correction messages are not yet known.

***The Use of Color and Brightness in Conspiracy Theories and Debunking Messages***

Analyzing visual framing of conspiracy theories is important since conspiracy theories are vastly circulated on social media platforms exchanging visual messages, such as YouTube (Hussain, Tokdemir, Agarwal, & Al-Khateeb, 2018) or Instagram (Vraga, Kim, Cook, & Bode, 2020). However, few studies examine visual frames of conspiracy theories (Brennen, Simon, & Nielsen, 2020). For correction messages, researchers suggest that visuals in correction messages help to overcome the effects of misinformation (King & Lazard, 2020). Compared to the emphasized importance of visuals in correction messages in experimental studies, how correction messages on social media platforms use visual frames to counterargue conspiracy theories is unknown. Analyzing visual frames debunking conspiracy theories on social media platforms is important since the examination allows us to understand how social media users employ visual frames to combat misinformation.

Especially, considering the social media environment becoming a contested ground between the misinformation spreaders and corrective crowd debunking misinformation, it is crucial to understand distinct visual tactics used both by conspirators and debunkers (Jung et al., 2020; Schmitt et al., 2018). Therefore, the present paper aims to fill in the gap between visual framing literature and misinformation and correction studies by examining visual frames employed in conspiracy theories and correction messages on social media platforms. More specifically, the present paper examines whether conspiracy theories and correction messages use visual modalities such as degree of color or saturation differently to persuade audiences.

One key difference that distinguishes conspiracy theories from correction messages is the strong association between conspiracy theories and the culture of paranoia (Aupers, 2012). Paranoia occurs by individuals' persecutory belief that harm will occur, and the harm is intended by others (Raihani & Bell, 2019). Conspiracy theories use framing devices to make audiences paranoid by demonizing established institutions and making audiences doubt factual knowledge produced by those institutions (Neville-Shepard, 2018). In contrast, correction messages try to rebuild epistemic authorities of established institutions (Kienhues, Jucks, & Bromme, 2020) and therefore are far from the culture of paranoia.

Another genre that uses framing devices to make audiences paranoid is the genre of horror (Pinedo, 2004). Like conspiracy theories, horror films also make audiences perceive that harm will occur and demonize one of the trustful characters (Pinedo, 2004). Therefore, by drawing from the color rhetoric commonly utilized by horror films to magnify paranoia from audiences, we can examine distinguishable visual attributes of conspiracy theories from correction messages. Compared to the lack of studies on visual framing in conspiracy theories, there are many computational works focusing on analyzing horror films' visual modality that contributes to create an atmosphere of paranoia (Rasheed et al., 2005).

Computational studies classifying films based on examining visual attributes demonstrated that horror films are more likely to use a low-lighting key and have lower variance in color than other film genres (Rasheed et al., 2005). The reason is that a) horror films use a low-lighting key to create suspense, and b) horror films usually adopt darker colors than other films. Since conspiracy theories also often try to create an atmosphere of paranoia to persuade audiences, conspiracy theories might use low-key lighting and lower color variances to attract viewers. Compared to conspiracy theories, correction messages aim to lower uncertainty and fear (Garrett et al., 2013). Since conspiracy theories are more likely to share horror films' visual attributes than correction (i.e., debunking) messages, we raise two research questions:

***RQ1: Do conspiracy videos use more low-key lighting than correction videos?***
***RQ2: Do conspiracy videos use lower color variances than correction videos?***

Conspiracy theories about climate change have existed for many years, and conspiracy theories related to COVID-19 are rapidly increasing after the outbreak of the disease (Mian & Khan, 2020). While conspiracy theories about climate change mostly build narratives to deny the threatening result of climate change (Douglas & Sutton, 2015), conspiracy theories about COVID-19 include both narratives denying the devastating consequence of COVID-19 and blaming the cause of COVID-19 to the government (e.g., the Chinese government) or technology (e.g., 5G technology) (Shahsavari, Holur, Wang, Tangherlini, & Roychowdhury, 2020).

However, some climate change conspiracy theories also suggest that climate change happens because of the government's nefarious intentions to harm citizens (e.g., geoengineering) (Allgaier, 2019). Since both conspiracy theories related to climate change and emerging conspiracy theories around COVID-19 have commonalities, visual frames of conspiracy theories around these issues might be similar. Therefore, we raise the third research question:

***RQ3*: *Do conspiracy videos about emerging science topics (COVID-19) and conspiracy videos about traditional science topics (Climate change) share the characteristics of horror films?***

### *A Computational Approach to Understand Visual Framing (in Conspiracy)*

There is a growing work among communication scholars that analyzes visual frames in media messages (Bucy & Joo, 2021) and applies computational skills to study visual frames (Joo & Steinert-Threlkeld, 2018; Peng, 2021). These computational methods are applied to 1) extract high-level features such as capturing facial expressions in the images or videos, and (or) 2) pull out visual modalities such as color or saturation in the images or videos.

The present paper uses computational methods to answer our research questions by 1) extracting frames from each video, 2) computing visual modalities from video frames, and 3) building and evaluating models to classify videos by image features and text features. In Figure 1, we presented a flowchart visualizing how we used computational techniques in each procedure to answer our research questions. This flowchart outlines our proposed computational method to analyze visual frames in communication research in a general form that can be applied to similar studies. We describe the considerations researchers need to consider while applying these methods in the following paragraphs.

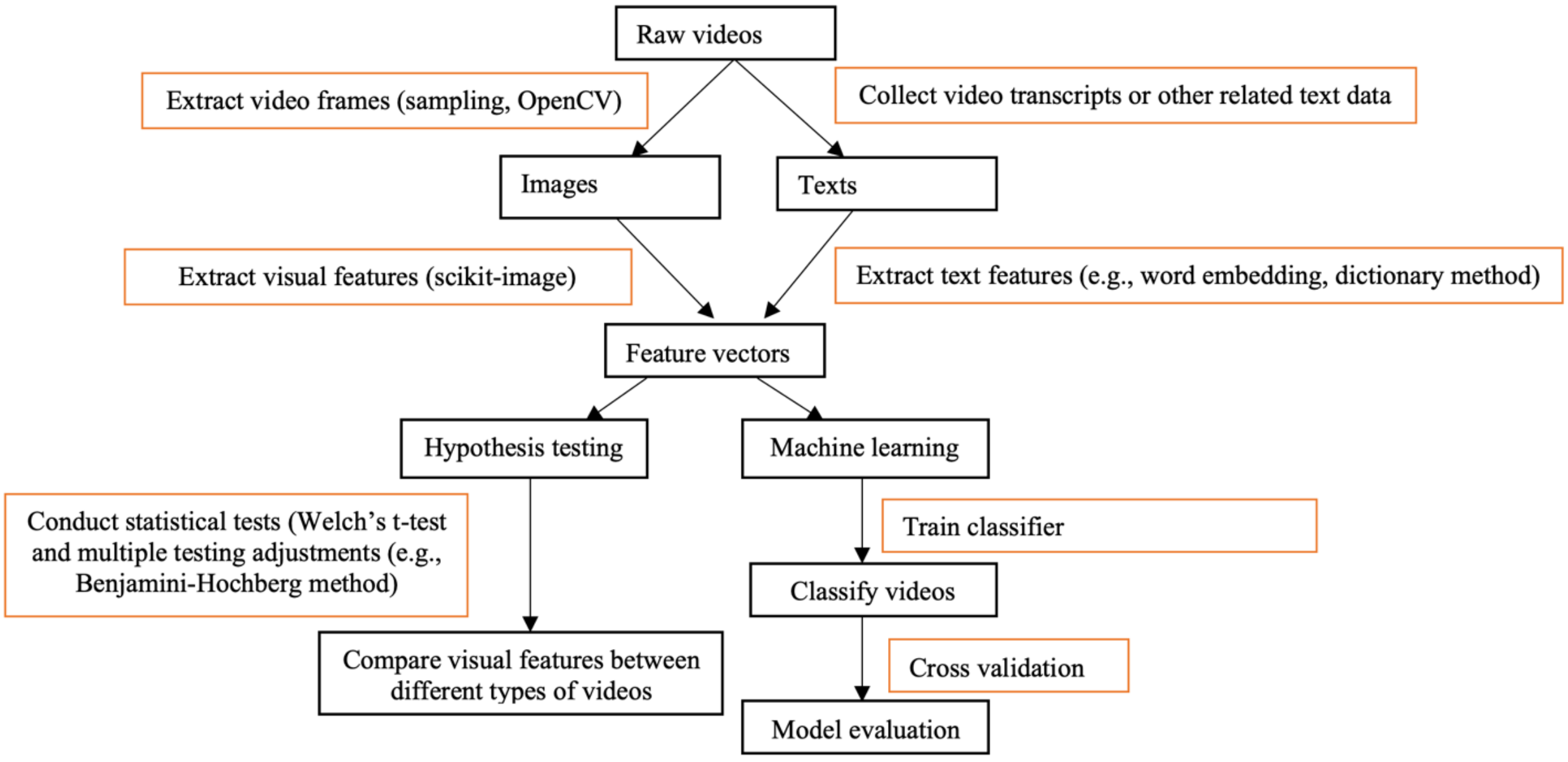

*Figure 1. Flowchart on how to analyze images from videos*

### ***Extracting frames from each video***

To extract visual frames from videos, researchers convert videos into an array of images by capturing frames (Derry et al., 2010). Converting videos into frames is an important first step if a researcher wants to process images computationally to extract visual feature information (Afifah, Nasrin, & Mukit, 2018). One of the computational tools that can be used by communication researchers to extract frames from videos is the OpenCV library (Culjak, Abram, Pribanic, Dzapo, & Cifrak, 2012). OpenCV is a comprehensive open-source computer vision library that provides application programming interfaces (APIs) for common programming languages such as C++, Python, Java, and MATLAB. In addition to providing state-of-the-art methods for image analysis, it also provides access to various image and video statistics, for example, the number of frames, frame rate, and frame size (Afifah et al., 2018).

***Computing visual modalities from video frames***

Communication studies employing computational techniques to analyze visual frames follow at least four approaches (Peng, 2021): 1) extracting attributes predetermined by open-source computer vision libraries and commercial APIs from videos or images (e.g., Peng, 2018), 2) using supervised machine learning models to extract feature attributes defined by researchers (e.g., Joo & Steinert-Threlkeld, 2018), 3) clustering features extracted from a supervised model pre-trained on a large image dataset and fine-tuned to the target dataset (i.e., transfer learning) (Peng, 2021), and 4) computationally analyzing visual modalities of images or videos, such as color or saturation (Peng & Jemmott, 2018). The present paper follows the last approach, which focuses on analyzing visual attributes of conspiracy and debunking videos.

After loading an image in OpenCV, researchers have access to its separate RGB ("red, green, blue") color channels, where RGB refers to the different hues of visible light that can be combined to produce different colors. Using a simple mathematical formula -- the conversion is implemented in OpenCV -- the RGB representation can be converted into the HSV format, which is an alternative color model that corresponds more closely to how humans perceive visual images (Joblove & Greenberg, 1978). HSV stands for "hue, saturation, value." Hue is used to capture the type of color or coloration. It ranges from 0 to 360 (expressed in degrees) corresponding to red (0-60), yellow (61-120), green (121-180), cyan (181-240), blue (241-300), and magenta (301-360). Saturation describes how colorful a given color is compared to how much gray it contains; its value range is 0-100 percent, corresponding to the amount of gray in a given color. Finally, "value" corresponds to the brightness of a color ranging from 0 to 100 percent, where 0 corresponds to black and 100 corresponds to the maximal expression of the color. While RGB and HSV color spaces are interconvertible, their corresponding channels captures different types of information about an image.

In addition to the RGB and HSV color schemes described in the previous paragraph, Peng and Jemmott (2018) also provide a computational method to calculate an image's colorfulness by following the formula from (Hasler & Suesstrunk, 2003). Peng and Jemmott (2018) estimate the colorfulness of the image based on each pixel's RGB values, which can be extracted from the OpenCV library. This colorfulness measure proposed by Hasler and Süsstrunk (2003) is computed via a linear combination of features from the CIELab color space (Hunt, 1989).

***Building models to classify conspiracy videos: image features vs. text features***

The present paper also introduces another computational approach to the communication field, that is, evaluating the importance of visual features in determining conspiracy videos. Researchers can use machine learning or deep learning techniques to assess if a combination of visual elements in images or videos can successfully distinguish images or videos with different characteristics (Birnbaum et al., 2020). For the present paper, we question whether visual attributes in YouTube videos can be utilized to distinguish between conspiracy videos and correction videos successfully. RQ1 and RQ2 focus on comparing each color aspect independently, whereas in RQ4, we study to what extent the combination of all the color aspects can be used to distinguish between conspiracy and debunking videos. More specifically, we examine if the information about low-key lighting and color variances can reliably differentiate between the two types of videos. Therefore, we raise the following research question:

***RQ4: Can information about low-key lighting and color variances be used to reliably distinguish between conspiracy and debunking videos?***

Researchers can utilize machine learning and deep learning models to predict whether videos classify as conspiracy from videos' visual features. Classical machine learning models such as random forests were developed with data in a tabular form in mind (Raschka, Patterson, & Nolet, 2020). In contrast, deep learning models are more attractive and effective for unstructured data, that is, image and text data in its original (raw) form prior to feature extraction (Raschka et al., 2020).

In particular, using performance estimates obtained via k-fold cross-validation, the focus is on analyzing feature modality (visual or textual) predominantly encodes information for predicting conspiracy videos. The present study compares classification models predicting conspiracy videos and debunking videos from 1) the textual features of the videos, 2) the visual features of the videos, and 3) both textual and visual features of the videos. By comparing these models, we try to answer the following research question:

***RQ5: Does the performance to distinguish between conspiracy and conspiracy-correction videos increase when we integrate textual and visual features of videos, compared to using one type alone?***

## Data and Method

### *Data Collection*

The main dataset we used in this paper consists of YouTube videos that are related to COVID-19 conspiracies. To sample COVID-19 conspiracy related videos, we drew upon search terms from literature and Google YouTube Search Trends (Au, Howard, & Bright, 2020; Pennycook, McPhetres, Zhang, Lu, & Rand, 2020). These search terms covered a variety of COVID-19 conspiracies from themes related to geopolitics (Wuhan Virus, Bioweapon), modern technology (5G conspiracy), and people's distrust of social elites (the Bill Gates population control conspiracy, Judy Mikovits, QAnon). We then used the YouTube API to sample the top 10 most viewed videos each day from March 1st to May 31st, 2020 (N=3668). The reason we sampled the top ten most viewed videos is because these are the videos is that these may have the highest impact on society. Among them, 2695 were English videos. The YouTube API provides access to video-related information such as title, description, channel-related information, and user click data (number of views, dislikes, likes, and comments). To collect the video transcripts, we used Pytube v. 9.6.0. Among the 2695 English videos, 1915 videos had transcripts available and 2154 videos had thumbnail images available, and these videos were used for analysis in this paper.

In addition to the COVID-19 conspiracy-related videos, to examine the image features beyond the COVID-19 topic to answer RQ3, we also collected conspiracy-related videos from the climate change topic as a robustness check. We used the search term "climate hacking" and "chemtrails" suggested by Allgaier (2019) as well as "climate denial" and "climate manipulation" to sample potential climate conspiracy related videos. Using the YouTube API, we collected the top-viewed videos that were published from 2015 to 2019 (n=277). After we collected the videos, we performed content analysis regarding whether a video is indeed about the conspiracy topic, and if so, whether the attitude of the host debunks the conspiracy or propagates conspiracies. We provide details of the content analysis method in the next section.

***Textual Content Analysis***

Since we are interested in comparing videos that propagate conspiracies vs debunk conspiracies, we developed two content variables to analyze the textual information (i.e., video transcripts) of each video: relevancy and attitude. *Relevancy* means whether a video is about the conspiracy topic. This variable serves as a sanity check to ensure that the videos we collected through the YouTube API are about the conspiracy topics we care about. *Attitude* means whether a video propagates conspiracies or debunks conspiracies. *Debunk* means a video refutes, disapproves, or (and) corrects a conspiracy. Two researchers manually coded a small sample of videos (n=60) and achieved inter-coder reliability of 93.5% for relevancy and 91.7% for attitude using Krippendorff's alpha (Krippendorff, 2011).

After achieving inter-coder agreement, 560 videos were hand labeled and served as the training dataset to train all the classifiers for our supervised machine learning method (for details, see Supplemental Material Appendix V). To train the textual classifier, we preprocessed the video transcripts by removing stop words, html tags, and punctuation, and converted the texts to lower case. To decrease the noise within the transcript texts, we generated chunked transcripts by extracting shorter transcript segments containing 40 words surrounding each conspiracy keyword (20 words before and 20 words after). In addition, to prepare the two sets of textual features for RQ4 and RQ5, we utilized dictionary methods and generated document embeddings from the transcripts as input features for the machine learning models. For the dictionary method, the textual features are represented by the number of unique association words appearing in the transcript with eight emotions (anger, fear, anticipation, trust, surprise, sadness, joy, and disgust) and two sentiments (negative and positive). We applied an existing emotion dictionary – NRC, which implements Saif Mohammad's NRC Emotion lexicon to code the transcript of each video (Mohammad & Turney, 2013). We also normalized the results by dividing the counts by the total number of words of the transcript to account for varying transcript lengths. For the second set of textual features, we used the doc2vec distributed memory method (Le & Mikolov, 2014) to convert each transcript into a feature vector. These features vectors are used in the model comparison part, which we discuss in more detail in a following section, "Building classification models to identify conspiracy videos."

Since the number of climate change videos was relatively small, a researcher hand labeled all the videos for the two textual variables. Among the 277 videos, 187 are relevant to climate change conspiracies and are not censored. Among them, 42% debunk conspiracies.

***Visual Content Analysis***

Our visual content analysis consists of the following steps: 1) extract frames from each video, 2) compute various color features for each frame and then aggregate at the video level, and 3) after achieving labels from the automatic annotation we explained in Textual Content Analysis above, we conducted the Welch two sample t-tests, controlled with the Benjamini-Hochberg procedure (1995), to examine whether the color statistics differ significantly between conspiracy and debunking videos.

To extract the frames from each video, we used the OpenCV Python package (v. 4.4), which allows researchers to extract frames from videos at a pre-determined rate. We extracted one frame per second. Among all the 2153 COVID-19 videos, the average number of frames is 1160. Among the COVID-19 videos, the minimum video length is 5 seconds, the maximum is 36,610 seconds, and the median is 482 seconds. In total, we analyzed 2,497,789 frames. For the climate change videos, the minimum video length is 6 seconds, the maximum is 9,561 seconds, the median is 490 seconds, and the average is 1012 seconds. In total, there are 243,939 frames we analyzed.

To calculate the color statistics of each frame, we used both OpenCV and the visual aesthetic GitHub repository from Peng and Jemmott (2018). We are interested in both the low-level features of color use (e.g., lighting) as well as the high-level features of color use (e.g., color variances). Using the two packages, we obtained RGB, HSV, brightness, contrast, and colorfulness statistics for each frame of a video. To compute the color statistics on the video level, we then calculated the median and variance of all the color features across all frames in a given video. The median measures the average degree of color use. The variance captures how a color feature varies across the frames of a video.

To address RQ1 and RQ2, which ask whether conspiracy videos differ significantly in the color and brightness use compared to correction videos, we performed Welch two sample t-tests comparing the means of the median and the variance distributions of all the color features between conspiracy and debunking videos. In addition, we utilized the Benjamini-Hochberg procedure which helps control the false discovery rate in multiple testing scenarios (Benjamini & Hochberg, 1995). To study the effect size of the differences of the groups we compared, we computed the Cohen's *d*, which estimates the effect size (i.e., the standardized difference of the population means).

***Building classification models to identify conspiracy videos***

To answer RQ4 and RQ5 which examine what features (including textual and visual) can best distinguish between conspiracy and conspiracy-correction videos, we trained multiple models on our hand-labeled COVID-19 dataset. In the first type of model, we only included the textual features from transcripts (n=560). To explore different options of textual features, we built three models on different textual feature sets, including a multi-layer perceptron with two densely connected layers of 64 hidden units trained on word frequencies from the whole transcript, a random forest model trained with the dictionary features, and a random forest model trained with document embedding features. In the second type of model, we only included the visual features (n=407, not censored and conspiracy relevant) to train the multi-layer perceptron and random forest models as before. In the third type of model, we included different combinations of textual and visual features to train the multilayer perceptron and random forest model: 1) transcript and visual features, 2) dictionary and visual features, and 3) document embedding and visual features.

Following common practices and recommendations when working with datasets of relatively small or moderate size, we evaluated the models using 10-fold cross-validation instead of creating separate training and test splits (Hawkins, Basak, & Mills, 2003). To compute the feature importance, we calculated the permutation importance of each of the feature to measure the weight of each feature using the Python eli5 v.0.10.1 package.

## Results

### *How Conspiracy and Debunking Videos Differ in their Color Use*

To address RQ1 and RQ2, whether conspiracy videos use more low-key lighting and lower color variances than correction videos, we compared the average color-related feature values (the values were averaged across all pixels and frames in each video) of 2153 conspiracy (n = 1677) and correction videos (n = 476), including hand-labeled and automatic annotated videos (See Supplemental Material Appendix IV). We found that conspiracy videos indeed exhibit a lower color saturation than correction videos (median saturation 70.87 versus 75.86, Figure 2). While this difference is statistically significant ($p = 0.02$), the saturation difference between conspiracy and debunking videos is relatively subtle (Cohen's $d = 0.121$) and may not be visible to a human observer. For other color features such as the RGB channel values, colorfulness, contrast, hue, and brightness, there is no statistically significant difference between all the frames in our conspiracy videos versus those in the correction videos.

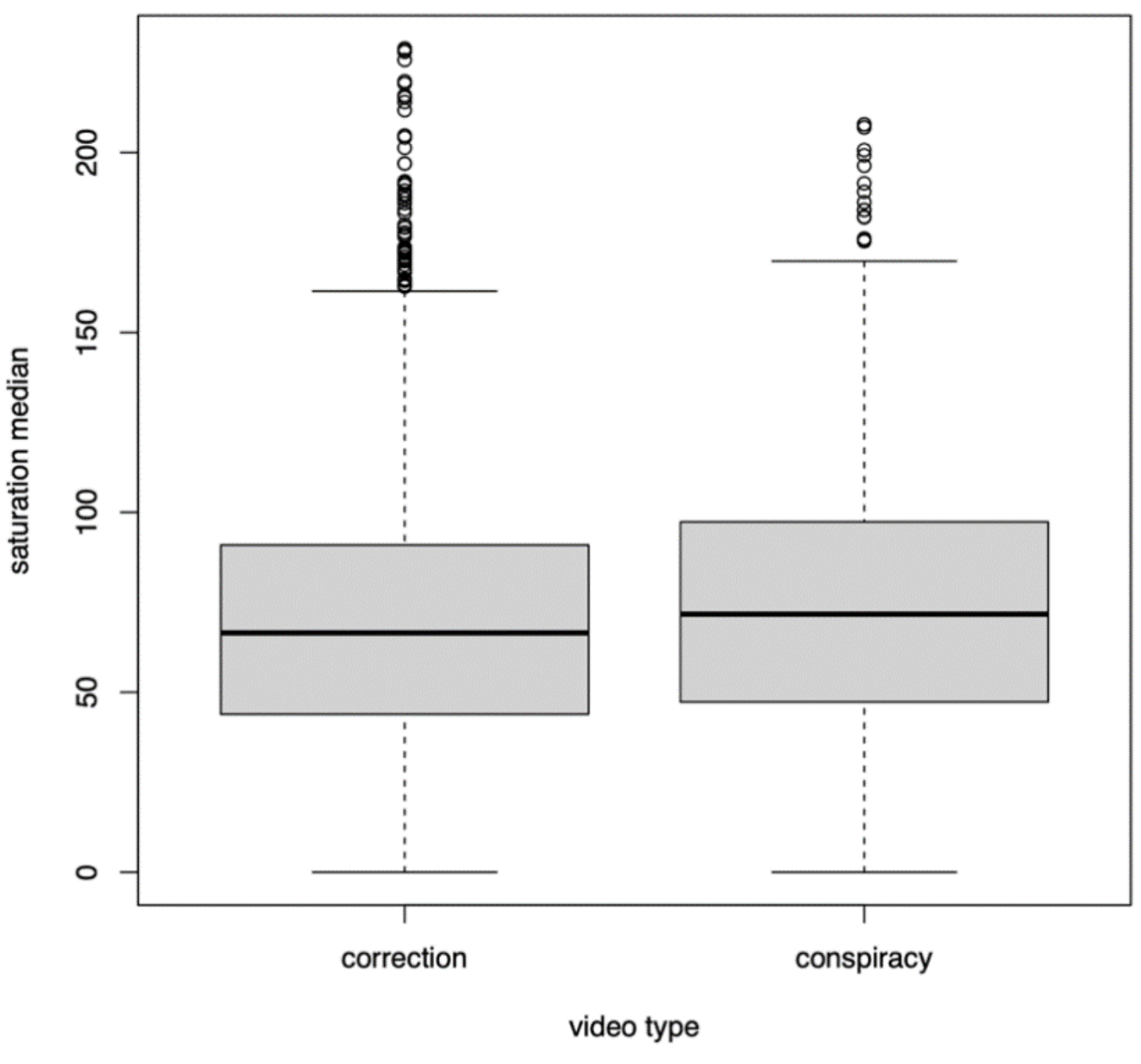

*Figure 2. Saturation comparison between conspiracy vs debunking videos*

However, notable differences were observed when we compared the frames from the first ten seconds of all conspiracy videos vs correction videos. The reason we conducted analyses on the first 10 seconds, which is considered as the average amount of time for a YouTube viewer to decide whether to engage with the content (McGavin, n.d.). Online creators need to catch viewers' eyes as soon as possible or they will lose the audience. Table 1 presents the median value of color features that are significant at the 0.05 level, when we compared frames using the first 10 seconds from all conspiracy videos and correction videos. We found that conspiracy videos use less red, green, and blue color compared to correction videos. We also found that conspiracy videos are less bright than debunking videos and used less color variance (contrast). The results show that effect size is small and thus color use in conspiracy vs correction videos differs significantly ($p < 0.05$) but not substantively (Cohen's $d < 0.5$).

Besides examining the first 10 seconds images of each video, we also studied thumbnails, a picture to represent the whole video and an important information for users to judge whether to watch the video. Similar to the color use of the first 10 seconds images, we found conspiracy videos used less brightness, contrast, and colorfulness, compared to correction videos and their differences are statistically significant (Table 2). We included in Supplemental Material Appendix VII examples of visual features for frames used in conspiracy propagation videos vs debunking videos.

Table 1. Significant results on COVID-19 conspiracy-related videos: using first 10 seconds

| Variable | Conspiracy videos | Correction videos | Uncorrected p-value | Benjamini–Hochberg (Q=0.05) | Cohen's *d* |
|---|---|---|---|---|---|
| RGB red (median) | 100.14 | 108.44 | 0.002 | 0.007 | 0.158 |
| Value (median) | 114.57 | 122.56 | 0.0035 | 0.014 | 0.151 |
| Brightness (median) | 96.53 | 103.63 | 0.007 | 0.021 | 0.141 |
| RGB green (median) | 94.56 | 101.25 | 0.013 | 0.029 | 0.130 |
| RGB blue (median) | 97.07 | 103.33 | 0.022 | 0.036 | 0.120 |
| Contrast (median) | 196.45 | 201.00 | 0.046 | 0.050 | 0.099 |

Table 2. Significant results on COVID-19 conspiracy-related videos: using thumbnails. BH = Benjamini-Hochberg (Q=0.05).

| | Variable | Conspiracy Videos | Correction Videos | Uncorrected p-value | BH | Cohen's *d* |
|---|---|---|---|---|---|---|
| Thumbnails | RGB Blue | 20.15 | 20.62 | 0.038 | 0.050 | 0.120 |
| | Saturation | 7.42 | 8.01 | 0.031 | 0.042 | 0.122 |
| | Value | 22.02 | 22.69 | 0.005 | 0.017 | 0.158 |
| | Brightness | 20.66 | 21.12 | 0.030 | 0.033 | 0.123 |
| | Contrast | 124.42 | 132.54 | 0.029 | 0.025 | 0.123 |
| | Colorfulness | 14.52 | 16.46 | 0.002 | 0.008 | 0.178 |

To examine whether our findings hold true going beyond the COVID-19 conspiracy, we also examined videos related to conspiracy theories from a traditional controversial science topic, climate change to answer RQ3. We repeated the same analyses as we did for COVID-19 videos: using all frames, the first 10 seconds frames, and the thumbnails. Similar to our findings about the COVID-19 conspiracy videos, we also found that climate change conspiracy videos used lower saturation, hue, and brightness. The effect size is larger compared to COVID-19 videos (e.g., Cohen's *d* for the median saturation is near 0.5, suggesting a medium effect size).

Table 3. Significant results on climate change conspiracy-related videos

| | **Variable** | **Conspiracy Videos** | **Correction Videos** | **Uncorrected p-value** | **Benjamini–Hochberg (Q=0.05)** | **Cohen's *d*** |
|---|---|---|---|---|---|---|
| Using all frames | No visual variable is significant | NA | NA | NA | NA | NA |
| Using first 10 seconds of frames | Saturation | 62.73 | 83.84 | 0.006 | 0.01 | 0.432 |
| | Hue | 55.30 | 71.10 | 0.011 | 0.02 | 0.398 |
| Thumbnails | Value | 93.74 | 106.83 | 0.037 | 0.025 | 0.312 |
| | RGB blue | 81.87 | 95.11 | 0.043 | 0.050 | 0.304 |

* For results of all the visual features variables (including not significant ones) described in Table 1-3, please see Supplemental Material Appendix I, II, and III.

### *Classifying Conspiracy Videos Comparing Visual vs Textual vs Integration Features*

To answer RQ4 and RQ5, which aim to explore what types of features are more useful in identifying conspiracy videos from correction videos, we built different types of models on the hand-labeled dataset, some using textual features only, some using visual features only, while others integrating both features. We used both a neural network (multilayer perceptron) and a random forest to compare the performance (e.g., precision and recall from 10-fold cross-validation) across models trained with different types of features. Table 4 summarizes the model performance. First, we found that using textual or visual features or an integration produce satisfying performance for identifying conspiracy videos. For instance, our model 1, 4 and 6 achieved the precision around 80% and recall around 90%. Using model 1 to classify our 1915 videos that have transcripts, we found that 84.46% propagated conspiracies, and 15.54% debunked conspiracies.

Second, in terms of whether data integration can achieve a better performance, we found inconclusive evidence. For instance, taking model 3, 4, and 5 for example, we found that although integrating textual features (e.g., emotion scores from analyzing video transcripts) and visual features (model 5) gives us the highest recall, the precision rate is lowest compared to only using the textual (model 3) or the visual features (model 4).

Table 4. Performance comparison in identifying conspiracy videos from correction videos. Precision and recall values incl. 95% confidence intervals. MLP = multi-layer perceptron; RF = random forest.

| | **Model** | **Precision** | **Recall** | **Feature** |
|---|---|---|---|---|
| 1 | MLP with transcript | 79.82 +/- 0.54 | 92.26 +/- 0.7 | word count (10000) |
| | MLP with visual | 74.08 +/- 0.77 | 75.95 +/- 7.79 | visual (20) |
| 2 | MLP with transcript + visual | 70.78 +/- 0.78 | 77.13 +/- 5.29 | word count (10000) + visual (20) |
| 3 | RF with 10 emotions | 69.72 +/- 0.33 | 88.19 +/- 1.57 | 10 emotions |
| 4 | RF with visual | 74.88 +/- 0.05 | 94.75 +/- 0.32 | 20 visuals |
| 5 | RF with 10 emotions + visual | 69.50 +/- 0.27 | 92.05 +/- 0.73 | 10 emotions + 20 visuals |
| 6 | RF with word embedding | 73.67 +/- 0.3 | 98.27 +/- 0.45 | word embedding (200 feature vector) |
| 7 | RF with word embedding + visual | 75.22 +/- 0.34 | 92.11 +/- 0.61 | word embedding (200 feature vector) + visuals (20) |

There could be several possible reasons why integration models like model 5 (in Table 4) do not necessarily increase model performance. We speculate that the combination of visual and textual features is redundant and thus an integrated approach does not improve model performance. Because both the textual features and the visual features result in good model prediction performance when used individually, the low precision score from the integration model (model 5) could be due to the noise in our small dataset. An alternative explanation is that a more sophisticated integration method, compared to our current integration of using simple combinations of emotion + visual features or word embedding + visual features, is required for boosting the performance of a model. We discuss the implication in the discussion section.

## Discussion and Future Work

### *How Conspiracy Videos Manipulate Color and Brightness*

Our paper advances the understanding of a core topic in communication -- visual framing in conspiracy theories. Conspiracy theories and correction messages co-exist and compete on social media. Despite of many studies that examine the textual features of conspiracies and how conspiracies diffuse on social media (Tingley & Wagner, 2017; Wood, 2018), little is understood about the visual features of conspiracies. Even less is understood about how visual features differ between conspiracies and correction messages. Studying visual framing of conspiracy is vital for advancing our understanding of conspiracy and misinformation in general as visuals serve as critical cues to influence people's emotion and behavior (Vraga et al., 2020).

In this paper, we focus on one aspect of visual framing: the use of visual modalities, to investigate how conspiracy videos manipulate visual modalities to communicate complex science issues from the recent COVID-19 conspiracies to the traditionally controversial climate change conspiracies. There are two noteworthy findings from our computational analysis on the images from YouTube videos on conspiracies and debunking conspiracies. First, in terms of the COVID-19 conspiracy, we found that conspiracy videos use lower color variance and brightness compared to debunking videos in general. However, the effect size in terms of this difference is small. This significant but not substantive difference poses challenges for platforms and researchers to use visuals to identity misinformation, as the difference in color and brightness does not differ substantively. It also poses challenges for science communicators to teach strategies to the public about how to distinguish conspiracy from truth. Despite the small visual difference between COVID-19 conspiracy videos vs debunking videos, the difference is much larger and appreciable for climate change conspiracies. The difference in the effect size when we compare different science topics suggest that visual framing could be contingent on the topic and thus, algorithms for identifying conspiracy through visuals need to tailor to the topic. Through revealing the complexity of visual use between conspiracy and debunking videos across science domains, our paper contributes to literature on visual framing, conspiracy, and misinformation correction on social media. It will be fruitful for future research to examine other visual aspects beyond color and brightness, and to also compare science topics with other topics such as health, and politics.

Interestingly, the color and brightness differences between conspiracy and debunking videos were only substantive and statistically significant for the video thumbnails and the first 10-second frames and not the full-length videos. Compared to correction video thumbnails, conspiracy video thumbnails shared more characteristics with horror films, and this was also true for the first 10-second frames of the videos. There could be two reasons. First, examining all frames could bring more noise to the data. For instance, some videos have a huge variation in the use of color features and even though we used the median color value to average across frames of a video, it still cannot capture the noise. The second reason, which is more likely, is that online creators deliberately orchestrate thumbnails and earlier frames of the video to catch attention of audiences (Zannettou, Chatzis, Papadamou, & Sirivianos, 2018). Since information provided by social media platforms is almost infinite, creating content that grabs attention is crucial in social media platforms (Webster, 2014). Thumbnails and first 10 seconds of frames of YouTube videos function as the clickbait ( Zannettou, Sirivianos, Blackburn, & Kourtellis, 2019) compared to the rest of frames in the videos, and therefore are more likely to use distinguishable textual and visual features to be "watched" by YouTube audiences. Color and brightness could be one effective way conspiracy video creators manipulate to attract their audience.

Understanding this difference in color and brightness between conspiracy videos and debunking videos can have important implications for future studies, since this difference contributes to understanding the visual rhetorical elements (i.e., lower color variance, darkness) added to conspiracy videos that further persuades audiences to believe in false knowledge. Recognizing this color rhetoric may contribute to the efforts of combating the propagation of conspiracy videos by detecting and flagging them, and by helping debunking video designers to think about the best way to counter the color rhetoric of conspiracy theories when encouraging audiences to correct their misperceptions about science.

### *Beyond Classification -- Gathering Insights from Small and Large Data Using Machine Learning*

From an image analysis perspective, our article offers several lessons for how machine learning can be used to extract insights from a dataset. In particular, our paper demonstrates how we can use machine learning to identify which combination of features can play a dominant role in distinguishing between different types of videos. The inspection of individual visual features, such as the saturation levels shown in Figure 2, is insufficient for distinguishing between conspiracy and debunking videos reliably. To get a better understanding, one might consider plotting multiple features at once; however, increasing the number of features to look at simultaneously increases the cognitive complexity and quickly becomes infeasible for humans.

Making sense out of datasets with multiple features is where machine learning can be utilized. After all, the objective of supervised machine learning algorithms employed in this study is to utilize the features in a dataset to maximize predictive performance. Hence, after training a highly accurate classifier, we can inspect how it utilizes the dataset's feature information to conclude about the important characteristics. For instance, we found that random forest models can use a combination of multiple visual features to make accurate predictions with a precision of approximately 75% (Table 4), where the three most important features hue, brightness, and contrast are approximately equally important to the model (see Table 5). Noting that hue measures the color content and is effectively uncorrelated to the brightness and contrast properties across the frames (see Supplemental Material Appendix IV) provides evidence that both color (RQ1) and brightness and low-key lighting (RQ2) are properties that can be used together (RQ4) to distinguish between conspiracy and correction videos.

When using machine learning for image analysis, two sets of methods may be considered. The first set consists of traditional machine learning methods designed for tabular (also known as "structured") datasets. Examples of these methods include random forests and multilayer perceptrons, which we employed in this study. A second set of methods includes deep neural network architectures designed for working with images or pixel inputs directly, which are known as "unstructured" data. Examples of this second set of methods include convolutional neural networks (Simonyan & Zisserman, 2015) and vision transformers (Khan et al., 2021). To make unstructured data suitable for traditional machine learning methods requires converting a video or image dataset into a tabular format. This is usually done by extracting features information via additional preprocessing steps, for instance, calculating the summary statistics of the RGB and HSV color channels and other summary features such as colorfulness (Hasler & Süsstrunk, 2003). In contrast, convolutional neural networks or vision transformers take the original images or video frames as input and perform feature extraction implicitly during model training. While this implicit feature extraction is attractive from a researcher's perspective, because a researcher does not have to spend time and resources extracting summary information from images, convolutional neural networks and vision transformers require substantially larger amounts of data compared to traditional machine learning models designed for tabular data (Figure 5).

In this project, we utilized traditional machine learning algorithms because the amount of labeled data (videos annotated as conspiracy or debunking) was limited to a few thousand videos. Based on other works concerned with training convolutional networks for video classification, we estimated that we would require at least hundreds of thousands of videos to train an accurate classifier (Karpathy et al., 2014). While traditional machine learning algorithms can also be applied to unstructured data (such as image frames extracted from videos) directly, the sizeable feature-to-sample ratio can be detrimental to performance due to the risk of overfitting. Hence, we proposed several manual feature extraction techniques, such as computing the color channel averages and variances as well as other color and brightness values. As a beneficial side effect, restricting the problem to a few well-defined features based on domain literature (e.g., communication) also simplifies the analysis of the model's features for making predictions. While it is easier to interpret the feature contributions from traditional machine learning methods trained on structured data, it is worth noting several methods have recently been developed to gain insights into features deep neural networks derive from unstructured data as well (Zhang & Zhu, 2018).

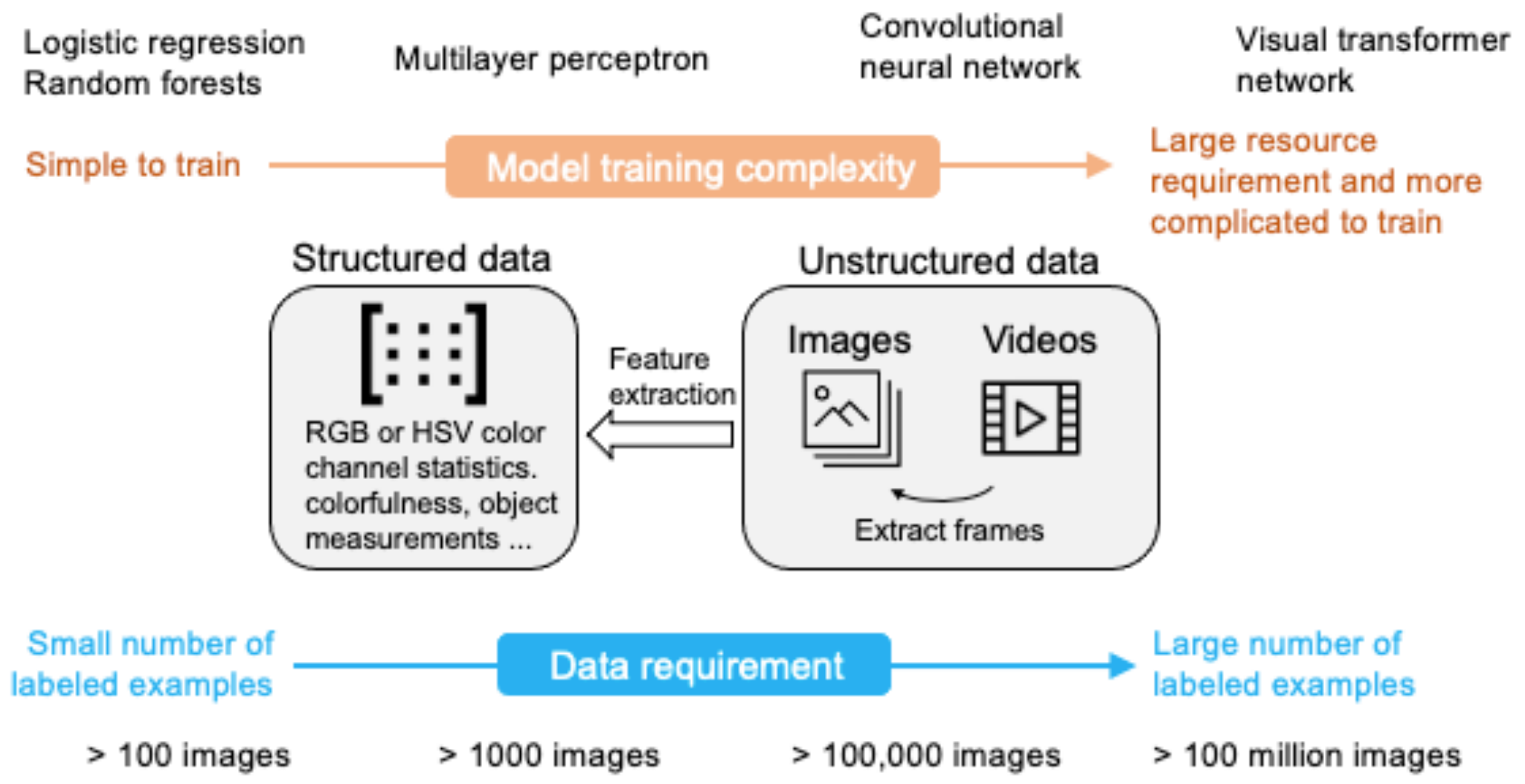

*Figure 5. Illustration of the complexity of the hyperparameter tuning tasks associated with different models (top) and recommended dataset sizes (bottom) when working with different models suited for structured and (or) unstructured data*

Our study shows that machine learning can yield concrete insights even if a dataset is small. Moreover, to approach a study in a time- and cost-effective manner, we recommend beginning with established building blocks such as those methods used in our computational analysis framework, before considering more complex and data-hungry deep learning models.

***Limitations and future directions***

We shall highlight that all models created in this study were developed to determine whether particular combinations of color and brightness relate to conspiracy content. The dataset the classifiers were trained on was specific to a random subset of COVID-19 videos (conspiracy videos and correction videos), and the resulting classifiers are thus not intended for application to general video databases such as YouTube. While this study does not focus on developing classifiers to flag conspiracy content, additional experiments show that classifiers trained on the same feature sets can also distinguish between COVID-19 conspiracy and general COVID-19 videos (Supplementary Material VI). This provides evidence that color and brightness features could serve a beneficial role when developing methods for automatically detecting conspiracy theory content online in future studies.

Furthermore, beyond gaining insights into the data, the methods described in this paper can be used as valuable performance baselines when collecting additional labeled data and experimenting with deep neural network architectures such as convolutional neural networks in future work. While deep neural networks trained on large datasets have predictive performance advantages over traditional methods, these methods come with increased hardware and tuning requirements (Sze, Chen, Yang, & Emer, 2017). It is also unclear whether intuitive, human-interpretable information about color and brightness for distinguishing different video categories can be extracted from deep neural networks trained on unstructured data. Future work may explore whether additional insights (for example, information about important image locations) via gradient-based localization methods can be gained from extending this study to using large video datasets. Lastly, a new research area called self-supervised learning has recently emerged, which allows researchers to train large deep learning models even though labeled data is scarce (Jing & Tian, 2020) opening new opportunities for applications of deep learning for studying visual manipulation in the digital era.

**Acknowledgements**

Kaiping Chen would like to thank the National Science Foundation for supporting this work under Grant No. 2027375. Additional support for this research was provided by the University of Wisconsin-Madison Office of the Vice Chancellor for Research and Graduate Education with funding from the Wisconsin Alumni Research Foundation.

## References

Au, H., Howard, P. N., & Bright, J. (2020, May 18). Social media misinformation on German Intelligence Reports. Retrieved from https://comprop.oii.ox.ac.uk/wp-content/uploads/sites/93/2020/06/ComProp-Coronavirus-Misinformation-Weekly-Briefing-18-05-2020.pdf

Afifah, F., Nasrin, S., & Mukit, A. (2018). Vehicle speed estimation using image processing. *Journal of Advanced Research in Applied Mechanics, 48*(1), 9-16.

Ahmed, W., Vidal-Alaball, J., Downing, J., & López Seguí, F. (2020). COVID-19 and the 5G Conspiracy Theory: Social Network Analysis of Twitter Data. *Journal of Medical Internet Research*, *22*(5), e19458. https://doi.org/10.2196/19458

Allgaier, J. (2019). Science and Environmental Communication on YouTube: Strategically distorted communications in online videos on climate change and climate engineering. *Frontiers in Communication*, *4*, 36. https://doi.org/10.3389/fcomm.2019.00036

Aupers, S. (2012). 'Trust no one': Modernization, paranoia and conspiracy culture. *European Journal of Communication*, *27*(1), 22–34. https://doi.org/10.1177/0267323111433566

Benjamini, Y., & Hochberg, Y. (1995). Controlling the False Discovery Rate: A Practical and Powerful Approach to Multiple Testing. *Journal of the Royal Statistical Society: Series B (Methodological)*, *57*(1), 289–300. https://doi.org/10.1111/j.2517-6161.1995.tb02031.x

Birnbaum, M. L., Norel, R., Van Meter, A., Ali, A. F., Arenare, E., Eyigoz, E., Agurto, C., Germano, N., Kane, J. M., & Cecchi, G. A. (2020). Identifying signals associated with psychiatric illness utilizing language and images posted to Facebook. *Npj Schizophrenia*, *6*(1), 38. https://doi.org/10.1038/s41537-020-00125-0

Brantner, C., Lobinger, K., & Wetzstein, I. (2011). Effects of visual framing on emotional responses and evaluations of news stories about the Gaza Conflict 2009. *Journalism & Mass Communication Quarterly*, *88*(3), 523–540. https://doi.org/10.1177/107769901108800304

Brennen, J. S., Simon, F., Howard, P. N., & Nielsen, R. K. (2020). Types, sources, and claims of COVID-19 misinformation. *Reuters Institute*, 7, 3-1.

Bucy, E. P., & Joo, J. (2021). Editors' Introduction: Visual politics, grand collaborative programs, and the opportunity to think big. *The International Journal of Press/Politics*, *26*(1), 5–21. https://doi.org/10.1177/1940161220970361

Cao, J., Qi, P., Yang, T., Guo, J., & Li, J. (2020). Exploring the role of visual content in gake news detection. *Disinformation, Misinformation, and Fake News in Social Media* (pp. 141-161).

Caviano, J.L., & López, M.A. (2010). How color rhetoric is used to persuade: Chromatic argumentation in visual statements. *Colour: Design & Creativity, 5*, 1-11.

Chong, D., & Druckman, J. N. (2007). Framing Theory. *Annual Review of Political Science*, *10*(1), 103–126. https://doi.org/10.1146/annurev.polisci.10.072805.103054

Culjak, I., Abram, D., Pribanic, T., Dzapo, H., & Cifrek, M. (2012, May). A brief introduction to OpenCV. *In 2012 proceedings of the 35th international convention MIPRO* (pp. 1725-1730). IEEE.

Derry, S. J., Pea, R. D., Barron, B., Engle, R. A., Erickson, F., Goldman, R., Hall, R., Koschmann, T., Lemke, J. L., Sherin, M. G., & Sherin, B. L. (2010). Conducting video research in the learning sciences: Guidance on selection, analysis, technology, and ethics. *Journal of the Learning Sciences*, *19*(1), 3–53. https://doi.org/10.1080/10508400903452884

Dixon, G. N., McKeever, B. W., Holton, A. E., Clarke, C., & Eosco, G. (2015). The power of a picture: Overcoming scientific misinformation by communicating weight-of-evidence information with visual exemplars: The power of a picture. *Journal of Communication*, *65*(4), 639–659. https://doi.org/10.1111/jcom.12159

Douglas, K. M., & Sutton, R. M. (2015). Climate change: Why the conspiracy theories are dangerous. *Bulletin of the Atomic Scientists*, *71*(2), 98–106. https://doi.org/10.1177/0096340215571908

DuBOSE, C. N., Cardello, A. V., & Maller, O. (1980). Effects of colorants and flavorants on identification, perceived flavor intensity, and hedonic quality of fruit-flavored beverages and cake. *Journal of Food Science*, *45*(5), 1393–1399. https://doi.org/10.1111/j.1365-2621.1980.tb06562.x

Filimonov, K., Russmann, U., & Svensson, J. (2016). Picturing the party: Instagram and party campaigning in the 2014 Swedish elections. *Social Media + Society*, *2*(3), 205630511666217. https://doi.org/10.1177/2056305116662179

Garber, L. L., Hyatt, E. M., & Starr, R. G. (2000). The effects of food color on perceived flavor. *Journal of Marketing Theory and Practice*, *8*(4), 59–72.

Garber, L. L., Jr, & Hyatt,  and E. M. (2003). Color as a tool for visual persuasion. In *Persuasive Imagery*. Routledge.

Garrett, R. K., Nisbet, E. C., & Lynch, E. K. (2013). Undermining the corrective effects of media-based political fact checking? The role of contextual cues and naïve theory: Undermining corrective effects. *Journal of Communication*, *63*(4), 617–637. https://doi.org/10.1111/jcom.12038

Geise, S. (2017). Visual framing. In P. Rössler (Ed), *The International Encyclopedia of Media Effects*, (pp. 1-12). NJ: John Wiley & Sons.

Gerend, M. A., & Sias, T. (2009). Message framing and color priming: How subtle threat cues affect persuasion. *Journal of Experimental Social Psychology*, *45*(4), 999–1002. https://doi.org/10.1016/j.jesp.2009.04.002

Halpern, D., Valenzuela, S., Katz, J., & Miranda, J. P. (2019). From belief in conspiracy theories to trust in others: Which factors influence exposure, believing and sharing fake news. In G. Meiselwitz (Ed.), *Social Computing and Social Media. Design, Human Behavior and Analytics* (Vol. 11578, pp. 217–232). Springer International Publishing. https://doi.org/10.1007/978-3-030-21902-4_16

Hasler, D., & Suesstrunk, S. E. (2003). *Measuring colorfulness in natural images* (B. E. Rogowitz & T. N. Pappas, Eds.; p. 87). https://doi.org/10.1117/12.477378

Hawkins, D. M., Basak, S. C., & Mills, D. (2003). Assessing model fit by cross-validation. *Journal of Chemical Information and Computer Sciences*, *43*(2), 579–586. https://doi.org/10.1021/ci025626i

Hine, T. (1996). *The Total Package*. NY: Little, Brown.

Hilbert, D. R. (1987). Color and color perception: A study in anthropocentric realism. Stanford University: Center for the Study of Language and Information.

Hollway, W., & Jefferson, T. (2005). Panic and perjury: A psychosocial exploration of agency. *British Journal of Social Psychology*, *44*(2), 147–163. https://doi.org/10.1348/014466604X18983

Huang, H.-Y., Shih, W.-S., & Hsu, W. H. (2008). Film classification based on low-level visual effect features. *Journal of Electronic Imaging*, *17*(2), 023011. https://doi.org/10.1117/1.2907215

Hussain, M. N., Tokdemir, S., Agarwal, N., & Al-Khateeb, S. (2018). Analyzing disinformation and crowd manipulation tactics on YouTube. *2018 IEEE/ACM International Conference on Advances in Social Networks Analysis and Mining (ASONAM)*, 1092–1095. https://doi.org/10.1109/ASONAM.2018.8508766

Joblove, G. H., & Greenberg, D. (1978). Color spaces for computer graphics. *Proceedings of the 5th Annual Conference on Computer Graphics and Interactive Techniques - SIGGRAPH '78*, 20–25. https://doi.org/10.1145/800248.807362

Joo, J., & Steinert-Threlkeld, Z. C. (2018). Image as data: Automated visual content analysis for political science. arXiv preprint arXiv:1810.015

Jing, L., & Tian, Y. (2020). Self-supervised visual feature learning with deep neural networks: A survey. *IEEE Transactions on Pattern Analysis and Machine Intelligence.* https://doi.org/ 10.1109/TPAMI.2020.2992393.

Jung, A.-K., Ross, B., & Stieglitz, S. (2020). Caution: Rumors ahead—A case study on the debunking of false information on Twitter. *Big Data & Society*, *7*(2), 205395172098012. https://doi.org/10.1177/2053951720980127

Karpathy, A., Toderici, G., Shetty, S., Leung, T., Sukthankar, R., & Fei-Fei, L. (2014). Large-scale video classification with convolutional neural networks. *In Proceedings of the IEEE conference on Computer Vision and Pattern Recognition* (pp. 1725-1732).

Khan, S., Naseer, M., Hayat, M., Zamir, S. W., Khan, F. S., & Shah, M. (2021). Transformers in vision: A Survey. ArXiv:2101.01169 [Cs]. http://arxiv.org/abs/2101.01169

Kienhues, D., Jucks, R., & Bromme, R. (2020). Sealing the gateways for post-truthism: Reestablishing the epistemic authority of science. *Educational Psychologist*, *55*(3), 144–154. https://doi.org/10.1080/00461520.2020.1784012

King, A. J., & Lazard, A. J. (2020). Advancing visual health communication research to improve infodemic response. *Health Communication*, *35*(14), 1723–1728. https://doi.org/10.1080/10410236.2020.1838094

Koh, B., & Cui, F. (2020). Visual persuasion: An exploration of the relation between the visual attributes of thumbnails and the view-through of videos. *Available at SSRN 3611735*.

Kou, Y., Gui, X., Chen, Y., & Pine, K. (2017). Conspiracy talk on social media: Collective sensemaking during a public health crisis. *Proceedings of the ACM on Human-Computer Interaction*, *1*(CSCW), 1–21. https://doi.org/10.1145/3134696

Kress, G. R., & Van Leeuwen, T. (1996). *Reading images: The grammar of visual design.* NY: Psychology Press.

Le, Q., & Mikolov, T. (2014, June). Distributed representations of sentences and documents. In International conference on machine learning (pp. 1188-1196). PMLR.

McGavin, R. (n.d.). *How to use the first 10 seconds of your video to hook your audience and increase views*. Retrieved from https://www.videomaker.com/article/c05/18826-how-to-use-the-first-10-seconds-of-your-video-to-hook-your-audience-and-increase-views

Meier, B. P., D'Agostino, P. R., Elliot, A. J., Maier, M. A., & Wilkowski, B. M. (2012). Color in context: Psychological context moderates the influence of red on approach- and avoidance-motivated behavior. *PLoS ONE*, *7*(7), e40333. https://doi.org/10.1371/journal.pone.0040333

Mian, A., & Khan, S. (2020). Coronavirus: The spread of misinformation. *BMC Medicine*, *18*(1), 89. https://doi.org/10.1186/s12916-020-01556-3

Mohammad, S. M., & Turney, P. D. (2013). Crowdsourci ng a word-emotion association lexicon. *Computational Intelligence*, *29*(3), 436–465. https://doi.org/10.1111/j.1467-8640.2012.00460.x

Moore, A. (2018). Conspiracies, Conspiracy Theories and Democracy. *Political Studies Review*, *16*(1), 2–12. https://doi.org/10.1111/1478-9302.12102

Murthy, V. H. (2021). *Confronting health misinformation: The U.S. surgeon general's advisory on building a healthy information environment.* Washington: U.S. Dept. of Health, Education, and Welfare, Public Health Service.

Neville-Shepard, R. (2018). Paranoid style and subtextual form in modern conspiracy rhetoric. *Southern Communication Journal*, *83*(2), 119–132. https://doi.org/10.1080/1041794X.2017.1423106

Oxman, E. (2010). Sensing the image: Roland Barthes and the affect of the visual. *SubStance*, *39*(2), 71–90. https://doi.org/10.1353/sub.0.0083

Peng, Y. (2018). Same candidates, different faces: Uncovering media bias in visual portrayals of presidential candidates with computer vision. *Journal of Communication 68*(5), 920–941. doi: 10.1093/joc/jqy041.

Peng, Y. (2021). What makes politicians' Instagram posts popular? Analyzing social media strategies of candidates and office holders with computer vision. *The International Journal of Press/Politics*, *26*(1), 143–166. https://doi.org/10.1177/1940161220964769

Peng, Y., & Jemmott, J. B. (2018). Feast for the eyes: Effects of food perceptions and computer vision features on food photo popularity. *International Journal of Communication, 12*. 313-336.

Pennycook, G., McPhetres, J., Zhang, Y., Lu, J. G., & Rand, D. G. (2020). Fighting COVID-19 misinformation on social media: Experimental evidence for a scalable accuracy-nudge intervention. *Psychological Science*, 095679762093905. https://doi.org/10.1177/0956797620939054

Pinedo, I. C. (2004). Postmodern elements of the contemporary horror film. *The Horror Film*, 85-117.

Raihani, N. J., & Bell, V. (2019). An evolutionary perspective on paranoia. *Nature Human Behaviour*, *3*(2), 114–121. https://doi.org/10.1038/s41562-018-0495-0

Raschka, S., Patterson, J., & Nolet, C. (2020). Machine learning in Python: Main developments and technology trends in data science, machine learning, and artificial intelligence. *Information*, *11*(4), 193. https://doi.org/10.3390/info11040193

Rasheed, Z., Sheikh, Y., & Shah, M. (2005). On the use of computable features for film classification. *IEEE Transactions on Circuits and Systems for Video Technology*, *15*(1), 52–64. https://doi.org/10.1109/TCSVT.2004.839993

Rodriguez, L., & Dimitrova, D. V. (2011). The levels of visual framing. *Journal of Visual Literacy*, *30*(1), 48–65. https://doi.org/10.1080/23796529.2011.11674684

Scheufele, D. A., & Krause, N. M. (2019). Science audiences, misinformation, and fake news. *Proceedings of the National Academy of Sciences*, *116*(16), 7662–7669. https://doi.org/10.1073/pnas.1805871115

Schmitt, J. B., Rieger, D., Rutkowski, O., & Ernst, J. (2018). Counter-messages as prevention or promotion of extremism?! The potential role of YouTube. *Journal of Communication*, *68*(4), 780–808. https://doi.org/10.1093/joc/jqy029

Shahsavari, S., Holur, P., Wang, T., Tangherlini, T. R., & Roychowdhury, V. (2020). Conspiracy in the time of corona: Automatic detection of emerging COVID-19 conspiracy theories in social media and the news. *Journal of Computational Social Science*, *3*(2), 279–317. https://doi.org/10.1007/s42001-020-00086-5

Simonyan, K., & Zisserman, A. (2014). Very deep convolutional networks for large-scale image recognition. arXiv preprint arXiv:1409.1556.

Singh, V. K., Ghosh, I., & Sonagara, D. (2021). Detecting fake news stories via multimodal analysis. *Journal of the Association for Information Science and Technology*, *72*(1), 3–17. https://doi.org/10.1002/asi.24359

Stemmet, C. (2012). Trust no truth: an analysis of the visual translation styles in the conspiracy film (Doctoral dissertation, University of Pretoria).

Sze, V., Chen, Y. H., Yang, T. J., & Emer, J. S. (2017, November). Efficient processing of deep neural networks: A tutorial and survey. *Proceedings of the IEEE, 105*(12) (pp. 2295-2329).

Tingley, D., & Wagner, G. (2017). Solar geoengineering and the chemtrails conspiracy on social media. *Palgrave Communications*, *3*(1), 12. https://doi.org/10.1057/s41599-017-0014-3

Valdez, P., & Mehrabian, A. (1994). Effects of color on emotions. *Journal of Experimental Psychology: General*, *123*(4), 394–409. https://doi.org/10.1037/0096-3445.123.4.394

van Prooijen, J.-W., & Douglas, K. M. (2017). Conspiracy theories as part of history: The role of societal crisis situations. *Memory Studies*, *10*(3), 323–333. https://doi.org/10.1177/1750698017701615

Vraga, E. K., Kim, S. C., & Cook, J. (2019). Testing logic-based and humor-based corrections for science, health, and political misinformation on social media. *Journal of Broadcasting & Electronic Media*, *63*(3), 393–414. https://doi.org/10.1080/08838151.2019.1653102

Vraga, E. K., Kim, S. C., Cook, J., & Bode, L. (2020). Testing the effectiveness of correction placement and type on Instagram. *The International Journal of Press/Politics*, *25*(4), 632–652. https://doi.org/10.1177/1940161220919082

Webster, J. G. (2014). *The Marketplace of Attention: How Audiences Take Shape in a Digital Age*. MIT Press.

Wood, M. J. (2018). Propagating and debunking conspiracy theories on Twitter during the 2015–2016 Zika Virus outbreak. *Cyberpsychology, Behavior, and Social Networking*, *21*(8), 485–490. https://doi.org/10.1089/cyber.2017.0669

Zannettou, S., Chatzis, S., Papadamou, K., & Sirivianos, M. (2018). The Good, the bad and the bait: Detecting and characterizing clickbait on YouTube. *2018 IEEE Security and Privacy Workshops (SPW)*, 63–69. https://doi.org/10.1109/SPW.2018.00018

Zannettou, S., Sirivianos, M., Blackburn, J., & Kourtellis, N. (2019). The Web of false information: Rumors, fake News, hoaxes, clickbait, and various other shenanigans. *Journal of Data and Information Quality*, *11*(3), 1–37. https://doi.org/10.1145/3309699

Zhang, Q. S., & Zhu, S. C. (2018). Visual interpretability for deep learning: a survey. *Frontiers of Information Technology & Electronic Engineering, 19*(1), 27-39.

Visual Framing of Science Conspiracy Videos:
Integrating Machine Learning with Communication Theories to Study the Use of Color

**Supplemental Material**

**Appendix I. Table of t-test result for all image variables for COVID-19 videos**
*(N=2154, using all frames)*

| **Variable** | **Conspiracy videos** | **Correction videos** | **Uncorrected p-value** |
|---|---|---|---|
| var_RGB red | 1215.739 | 1185.369 | 0.6739 |
| var_RGB green | 1177.720 | 1179.066 | 0.9852 |
| var_RGB blue | 1189.838 | 1177.964 | 0.8705 |
| var_Hue | 423.8709 | 424.1010 | 0.9923 |
| var_Saturation | 976.9691 | 945.8499 | 0.6441 |
| var_Value | 1136.127 | 1114.266 | 0.7493 |
| var_Contrast | 1068.7221 | 999.0364 | 0.3323 |
| var_Colorfulness | 265.8281 | 251.5830 | 0.4086 |
| var_Brightness | 1133.952 | 1127.511 | 0.9266 |
| var_Brightness_sd | 175.9654 | 162.1243 | 0.1872 |
| median_ RGB red | 112.1830 | 112.1922 | 0.9969 |
| median_ RGB green | 106.4258 | 105.1905 | 0.6041 |
| median_ RGB blue | 107.316 | 106.623 | 0.7723 |
| median_Hue | 63.83538 | 62.55247 | 0.3845 |
| median_Saturation | 70.87274 | 75.86185 | 0.02258 |
| median_Value | 125.0447 | 125.4073 | 0.875 |
| median_Brightness | 108.2065 | 107.3038 | 0.698 |
| median_Brightness_sd | 62.08177 | 60.98694 | 0.1696 |
| median_Contrast | 207.3763 | 204.9023 | 0.1572 |
| median_Colorfulness | 38.26796 | 39.47644 | 0.2246 |
| RGB red_mean | 112.8669 | 113.7645 | 0.6656 |
| RGB green_mean | 107.3832 | 107.3881 | 0.9981 |
| RGB blue_mean | 108.3935 | 108.8993 | 0.811 |
| Hue_mean | 63.62565 | 62.39608 | 0.3463 |
| Saturation_mean | 72.79677 | 76.98192 | 0.03218 |
| Value_mean | 125.4011 | 126.8720 | 0.4708 |
| Brightness_mean | 109.1377 | 109.4675 | 0.8723 |
| Brightness_sd_mean | 62.04446 | 60.97108 | 0.1369 |
| Contrast_mean | 203.2594 | 201.8554 | 0.3825 |
| Colorfulness_mean | 39.63218 | 40.95035 | 0.1553 |
| Color_lag | 0.8174218 | 0.7998421 | 0.08832 |

**Appendix II. Table of t-test result for all image variables for COVID-19 videos** *(N=2154, using frames of the middle 10 seconds)*

| **Variable** | **Conspiracy Videos** | **Correction Videos** | **Uncorrected p-value** |
|---|---|---|---|
| var_RGB red | 365.3034 | 383.7097 | 0.7469 |
| var_RGB green | 357.2518 | 370.0541 | 0.8167 |
| var_RGB blue | 374.8691 | 407.4505 | 0.5863 |
| var_Hue | 136.5714 | 143.5362 | 0.7246 |
| var_Saturation | 306.9000 | 404.8419 | 0.133 |
| var_Value | 356.0932 | 341.4090 | 0.7747 |
| var_Contrast | 398.6923 | 413.4284 | 0.813 |
| var_Colorfulness | 87.51726 | 94.78349 | 0.6202 |
| var_Brightness | 344.8397 | 357.9431 | 0.8096 |
| var_Bright_sd | 60.35594 | 59.13547 | 0.8839 |
| median_ RGB red | 113.8212 | 113.0322 | 0.7675 |
| median_ RGB green | 108.3034 | 107.9242 | 0.8874 |
| median_ RGB blue | 109.1153 | 108.9333 | 0.9463 |
| median_Hue | 64.08969 | 61.81564 | 0.1577 |
| median_Satuation | 71.08050 | 75.24902 | 0.08426 |
| median_Value | 125.8474 | 126.4151 | 0.8281 |
| median_Brightness | 109.9978 | 109.5861 | 0.8755 |
| median_Brightness_sd | 62.18543 | 61.17101 | 0.2704 |
| median_Contrast | 204.6631 | 204.0021 | 0.7494 |
| median_Colorfulness | 38.66954 | 39.99781 | 0.2521 |
| RGB red_mean | 114.5861 | 113.8367 | 0.7702 |
| RGB green_mean | 109.0518 | 108.8335 | 0.9327 |
| RGB blue_mean | 109.8354 | 109.7700 | 0.9799 |
| Hue_mean | 63.97098 | 61.71626 | 0.1472 |
| Saturation_mean | 71.38565 | 75.55081 | 0.07209 |
| Value_mean | 126.5876 | 127.2317 | 0.7982 |
| Brightness_mean | 110.7955 | 110.4373 | 0.8876 |
| Brightness_sd_mean | 62.12893 | 61.35369 | 0.3776 |
| Contrast_mean | 203.9304 | 203.6604 | 0.8899 |
| Colorfulness_mean | 39.04694 | 40.30429 | 0.2635 |
| Color_lag | 0.3858484 | 0.4248635 | 0.04911 |

**Appendix III. Table of t-test result for all image variables for COVID-19 videos** *(N=2491, using thumbnails)*

| **Variable** | **Conspiracy Videos** | **Correction Videos** | **Uncorrected p-value** |
|---|---|---|---|
| RGB red | 20.88913 | 21.32993 | 0.05261 |
| RGB green | 20.25671 | 20.66547 | 0.05401 |
| RGB blue | 20.15250 | 20.62185 | 0.03824 |
| Hue | 4.884541 | 5.126327 | 0.05078 |
| Saturation | 7.422061 | 8.007414 | 0.03093 |
| Value | 22.02074 | 22.69314 | 0.005466 |
| Brightness | 20.65670 | 21.11934 | 0.02985 |
| Contrast | 124.4170 | 132.5391 | 0.02876 |
| Colorfulness | 14.52372 | 16.46495 | 0.002103 |

## Appendix IV. Correlation matrix of the visual features used in Model 4

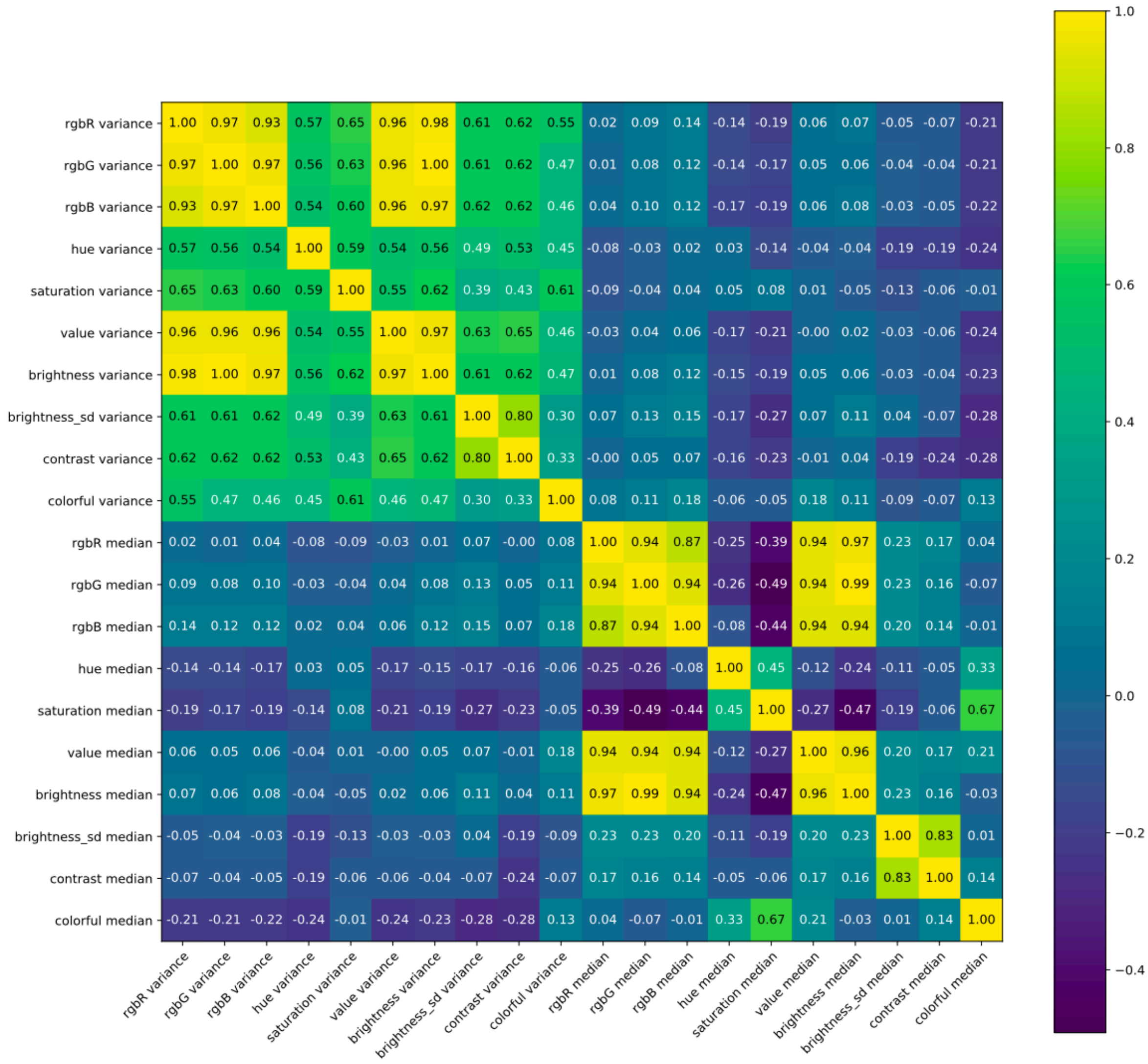

## Appendix V. Details of hand labeling process and the use of SML for automatic annotation

First, to acquire a reliable training dataset, we conducted an inter-coder agreement check between two coders to ensure that the two coders are reliable and consistent in judging whether the attitude of the video is about conspiracy propagation or debunking. During this inter-coder agreement checking process, multiple rounds of discussions happened to clarify the coding rules.

Finally, two coders coded a sample of 60 videos to perform an intercoder reliability check. The coders achieved 91.7% using Krippendorff's alpha for the content variable *Attitude*, indicating a high intercoder agreement. After this inter-coder agreement was achieved, one researcher hand-annotated the remaining 560 videos that constitute our training dataset.

After the training dataset was annotated by human researchers, we then performed data cleaning on the video transcripts by removing stop words, html tags, punctuation, and converting the texts to lower case. To use the text as input for a machine learning model, we mapped the texts' words into one-hot encoded vectors, which served as input to a multi-layer perceptron with 2 densely connected layers of 64 hidden elements using Python's scikit-learn package. This classifier was trained to predict attitude assigned by the coders within the processed training dataset. Under 10-fold cross-validation, this approach achieved 86.48% precision and 97.19% recall for our content analysis variable *Relevancy* and 79.82% precision and 92.26% recall for our content analysis variable *Attitude*.

Finally, after cross-validation, a multilayer perceptron model built on the transcripts of the whole training set was applied to the entire corpus of the collected English videos to predict the *Relevancy* and *Attitude* for those conspiracy-relevant videos.

## Appendix VI. New baseline analysis for distinguishing conspiracy videos from normal COVID19 videos

In our main manuscript, we answered RQ4-5 focusing on studying conspiracy propagation videos and debunking videos. However, it is also valuable to examine how our visual and textual features can distinguish conspiracy propagation videos from the normal COVID19 videos. This new baseline analysis is valuable as a first step toward developing mechanisms to identify conspiracy videos about COVID19 and other topics on social media. Therefore, in this section, we present this new baseline analysis.

### Data Collection

We conducted the normal COVID-19 video collection using the search term "COVID19" through the YouTube API. For each day between March 1st – May 31st, 2020, we obtained the top ten most viewed videos associated with the search term. A total of 514 normal COVID19 videos were collected, in the size comparable to the number of debunking videos analyzed in our main manuscript.

Next, we checked whether these videos based on the search term COVID19 were indeed normal COVID19 videos. We selected a random sample of 50 videos, and two researchers watched each video to judge whether it was "normal" (neither conspiracy or debunking content). Only 1 video talked about conspiracy-related theories, which we discarded from the dataset. The remaining videos talked about normal COVID19 topics such as giving health information about facial mask wearing, or health tips about how to cope with the COVID-19 pandemic. Finally, we combined the normal COVID-19 videos and conspiracy videos to form the dataset for this baseline analysis.

To prepare our data for analysis, we filtered out the English videos using CLD2 via the PYCLD2 Python package (https://pypi.org/project/pycld2/) and we also collected the transcript for all the videos using Pytube (https://pytube.io/en/latest/). In addition, we post-processed the video transcript for each of the video by reformatting the timestamp and removing useless information (e.g., html, tags, etc.). Furthermore, we applied existing emotion dictionaries – NRC, which implements Saif Mohammad's NRC Emotion lexicon (https://saifmohammad.com/WebPages/NRC-Emotion-Lexicon.htm) to code the transcript of each video in terms of each emotion dimension (e.g., fear, joy). In addition, we summed up all the emotions to obtain one aggregated variable called *total emotion words*. We also normalized the results by dividing the counts by the total number of characters of the transcript, since the length of each transcript varies. For the visual statistics, we first collected all the MP4s using Pytube and extracted their frames every second using OpenCV (https://opencv.org). Then, we calculated the average and variance of colorfulness, brightness, contrast, hue, saturation, value, and RGB red, RGB green, RGB blue for each video.

**Modeling Process**

The extracted textual and visual features, as described in the *Data Collection* section, were used to build the models. The model setups are the same as our main manuscript's analysis for RQ4-5.

**Result**

Generally, the model performance for classifying COVID19 conspiracy propagation videos versus normal videos was better than the model performance for classifying conspiracy versus debunking videos, as it can be seen when comparing models 1,3 and 6 in Table S2 to the models in Table 4 in the main manuscript. This may be because normal COVID19 videos have more distinguishable styles than conspiracy videos. In addition, for this new baseline analysis, textual features seem to be a discriminatory variable than visual features, as it can be seen from comparing model 1 and 2 in Table S2.

Table S2. Performance comparison in distinguishing conspiracy videos from normal COVID19 videos. Precision and recall values incl. 95% confidence intervals. MLP = multi-layer perceptron; RF = random forest.

| | Model | Precision | Recall | Feature |
|---|---|---|---|---|
| 1 | MLP with transcript | 93.41 +/- 0.3 | 94.99 +/- 0.42 | Word count (10000) |
| | MLP with visual | 65.1 +/- 1.07 | 69.17 +/- 5.97 | visual (20) |
| 2 | MLP with transcript + visual | 72.29 +/- 0.79 | 70.72+/- 6.31 | Word count (10000) + visual (20) |
| 3 | RF with 10 emotions | 99.25 +/- 0.06 | 98.59 +/- 0.15 | 10 emotions |
| 4 | RF with visual | 74.77 +/- 0.34 | 81.83 +/- 0.23 | 20 visuals |
| 5 | RF with 10 emotions + visual | 99.21 +/- 0.06 | 98.78 +/- 0.08 | 10 emotions + 20 visuals |
| 6 | RF with word embedding | 85.84 +/- 0.52 | 96.92 +/- 0.36 | Word embedding (200 feature vector) |
| 7 | RF with word embedding + visual | 85.45 +/- 0.4 | 93.30 +/- 0.31 | Word embedding (200 feature vector) + visuals (20) |

**Appendix VII. Example of images that illustrate extremes of color features for conspiracy propagation vs debunking videos.**

| COVID Conspiracy Propagation | COVID Correction |
|---|---|
| 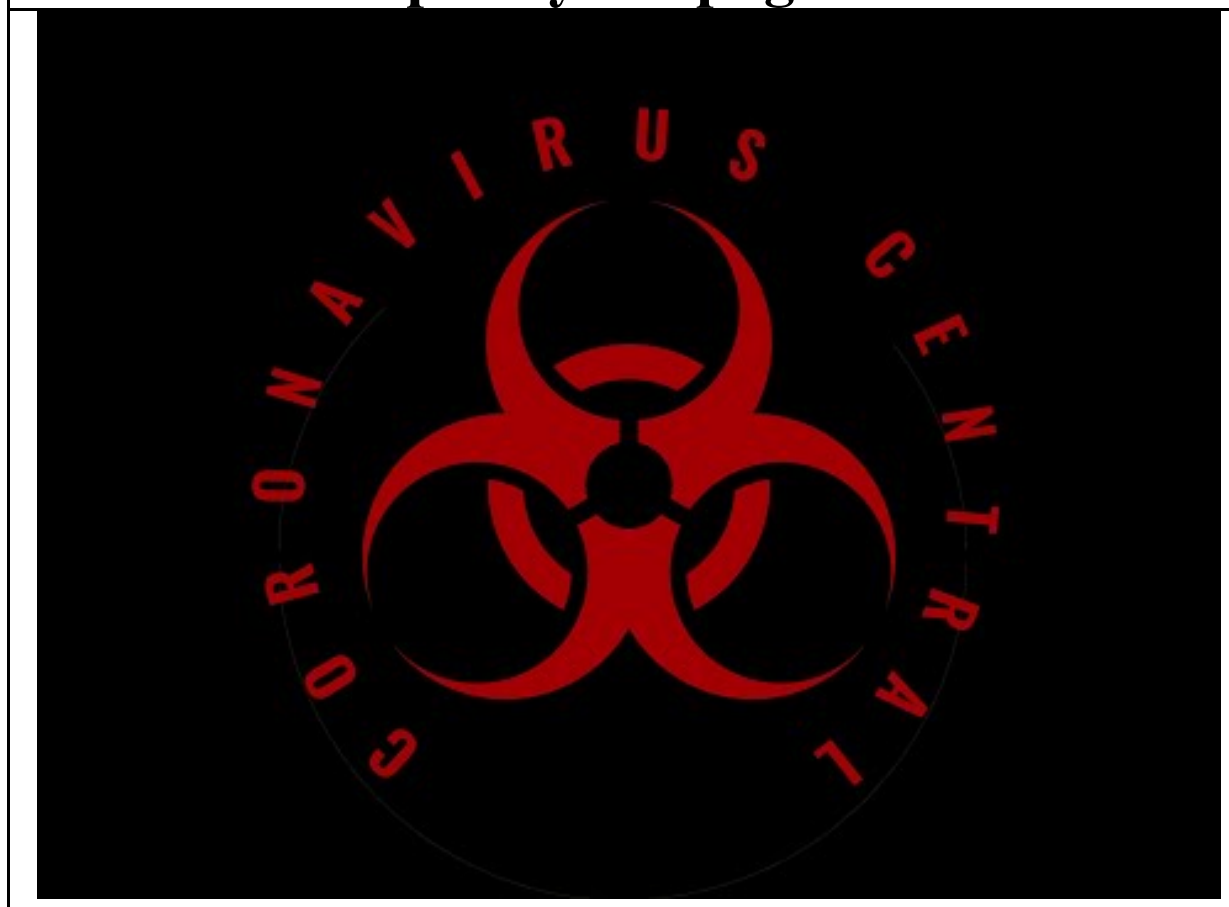<br>RGB Blue = 12.55938(low extreme)<br>https://i.ytimg.com/vi/pjI121Cmoko/hqdefault.jpg | 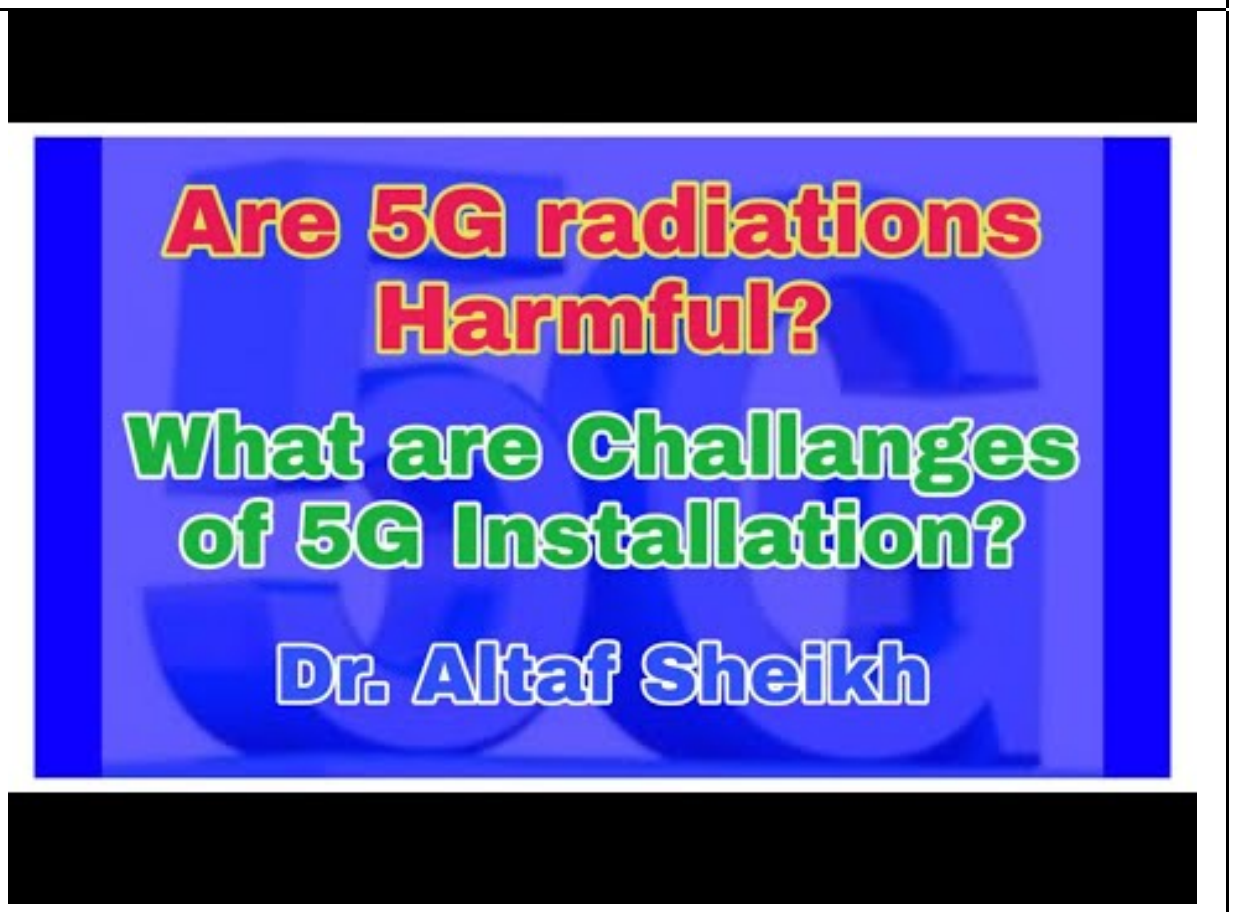<br>RGB blue = 32.17993 (high extreme)<br>https://i.ytimg.com/vi/Z4xVxBuP_wY/hqdefault.jpg |
| 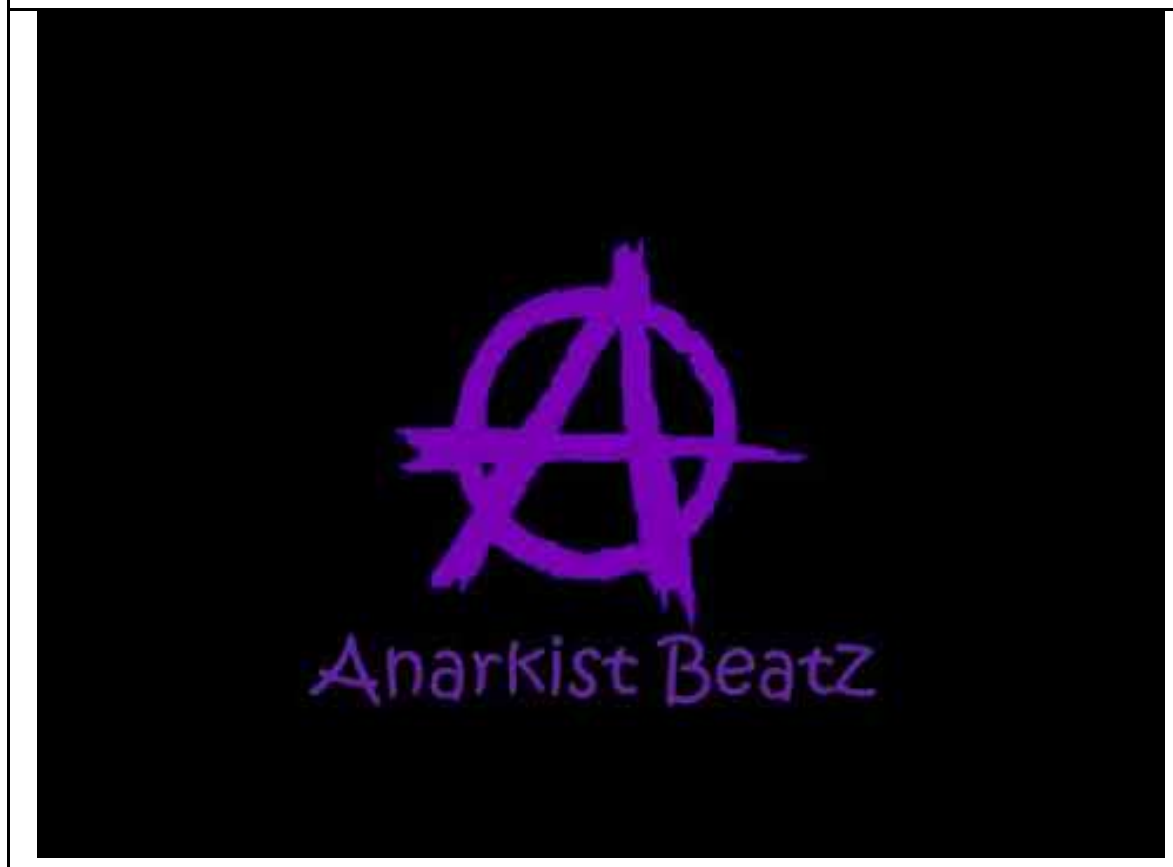<br>HSV value = 13.78679 (low extreme)<br>https://i.ytimg.com/vi/6EXD4IhHm8Y/hqdefault.jpg | 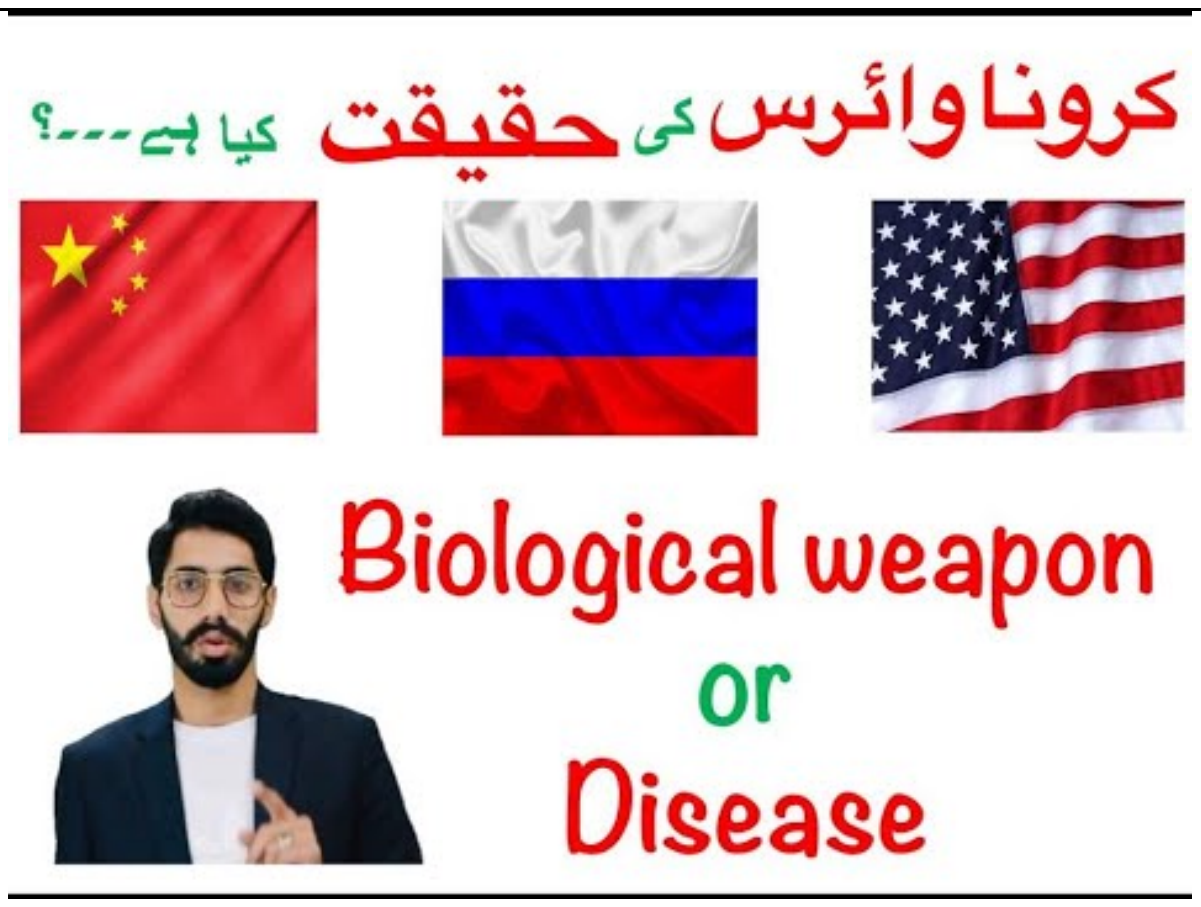<br>HSV value = 37.47461 (high extreme)<br>https://i.ytimg.com/vi/wHK7jbCQP48/hqdefault.jpg |

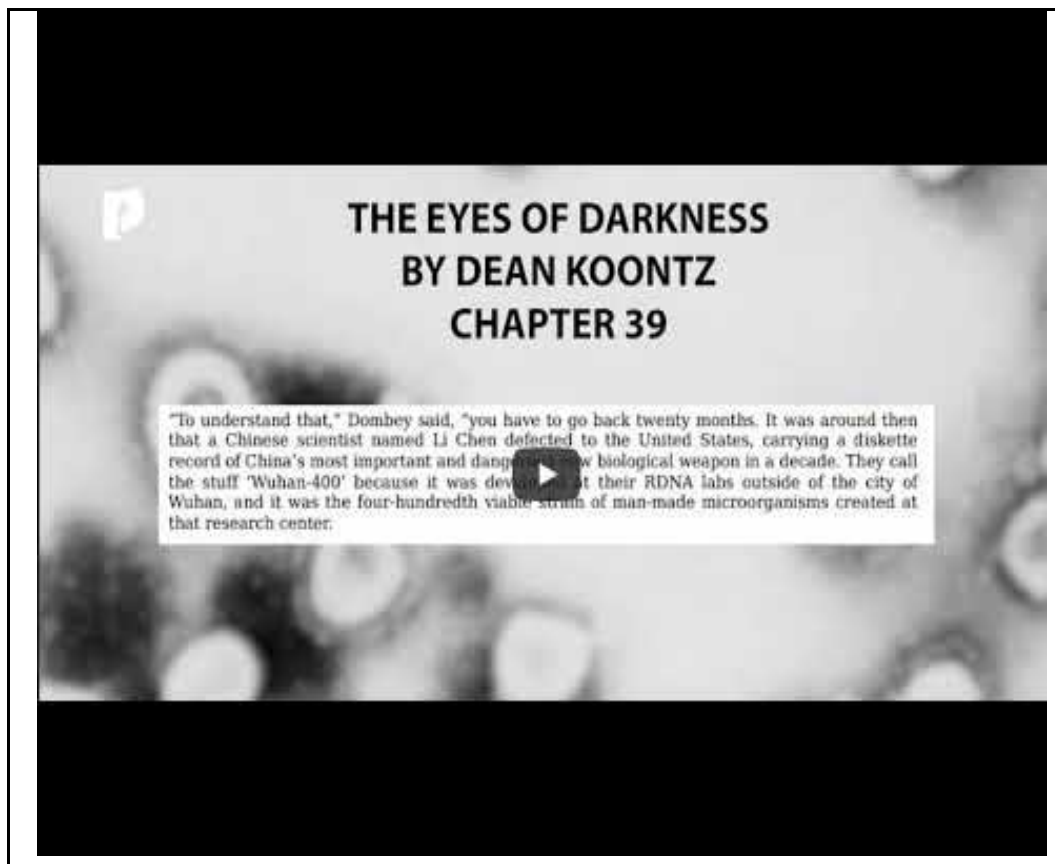

HSV saturation = 0.000962 (low extreme)
https://i.ytimg.com/vi/OvVgJHDUpTg/hqdefault.jpg

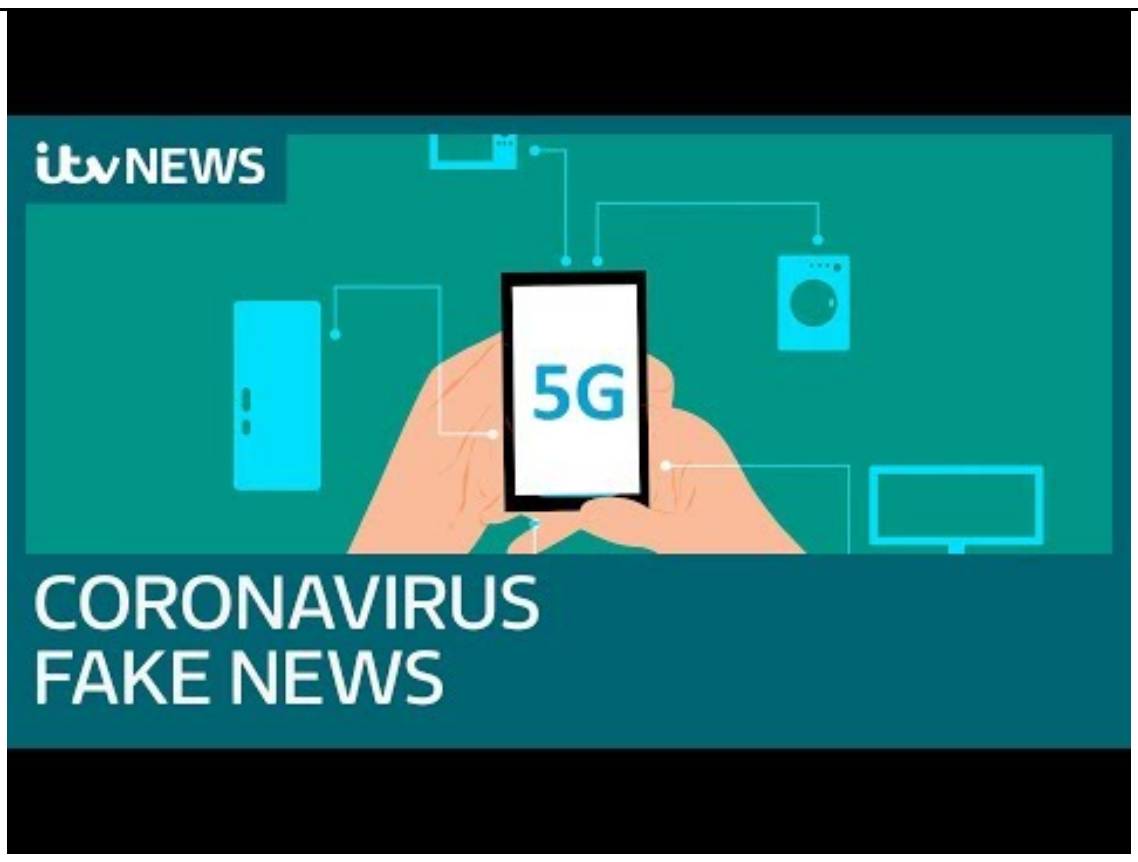

HSV saturation = 19.63112 (high extreme)
https://i.ytimg.com/vi/FrBk1xXvYzY/hqdefault.jpg

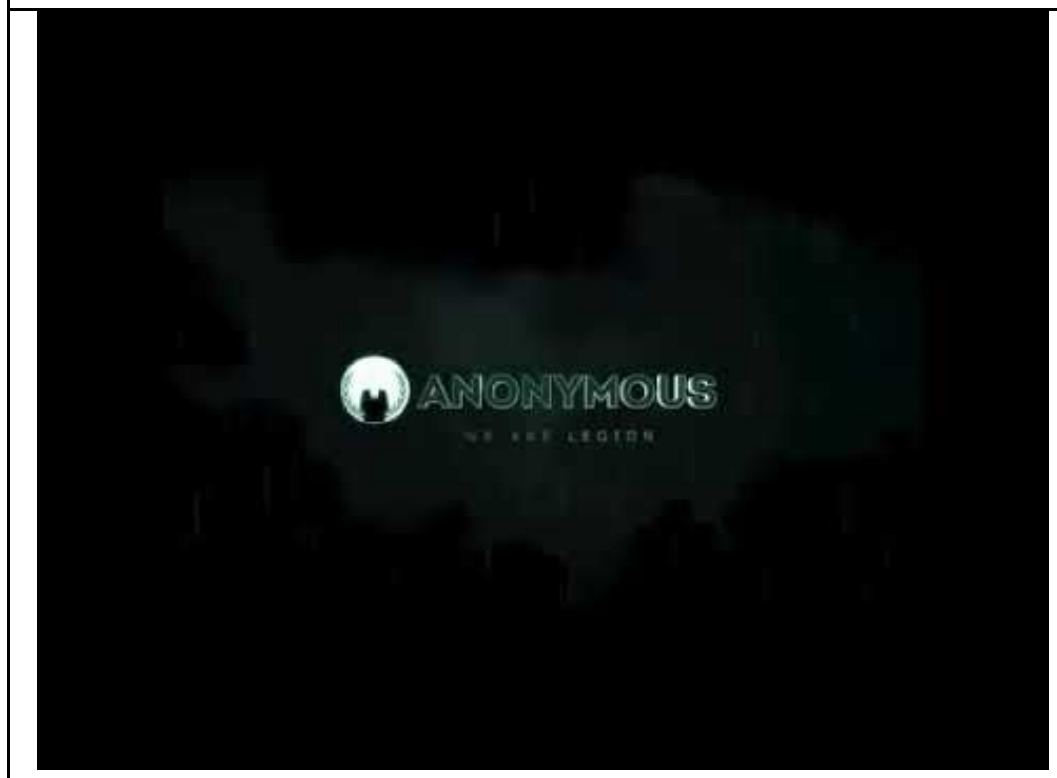

Contrast = 14 (low extreme)
https://i.ytimg.com/vi/iUNoiHrpIcw/hqdefault.jpg

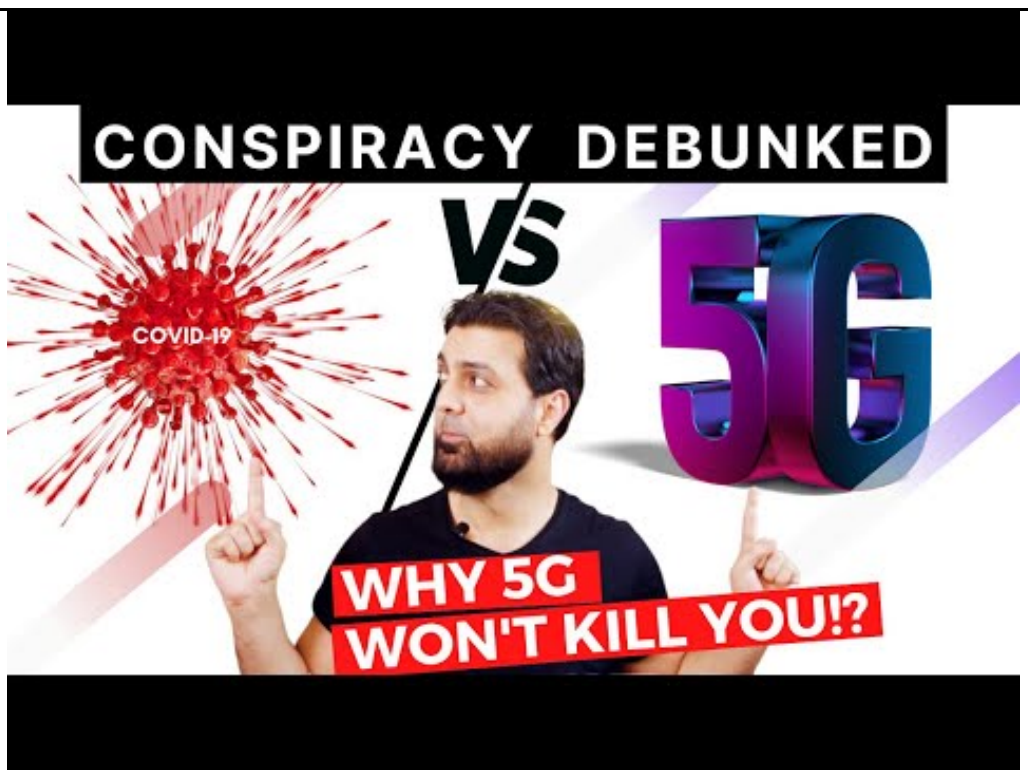

Contrast = 253 (high extreme)
https://i.ytimg.com/vi/Af1hJ_rifs8/hqdefault.jpg

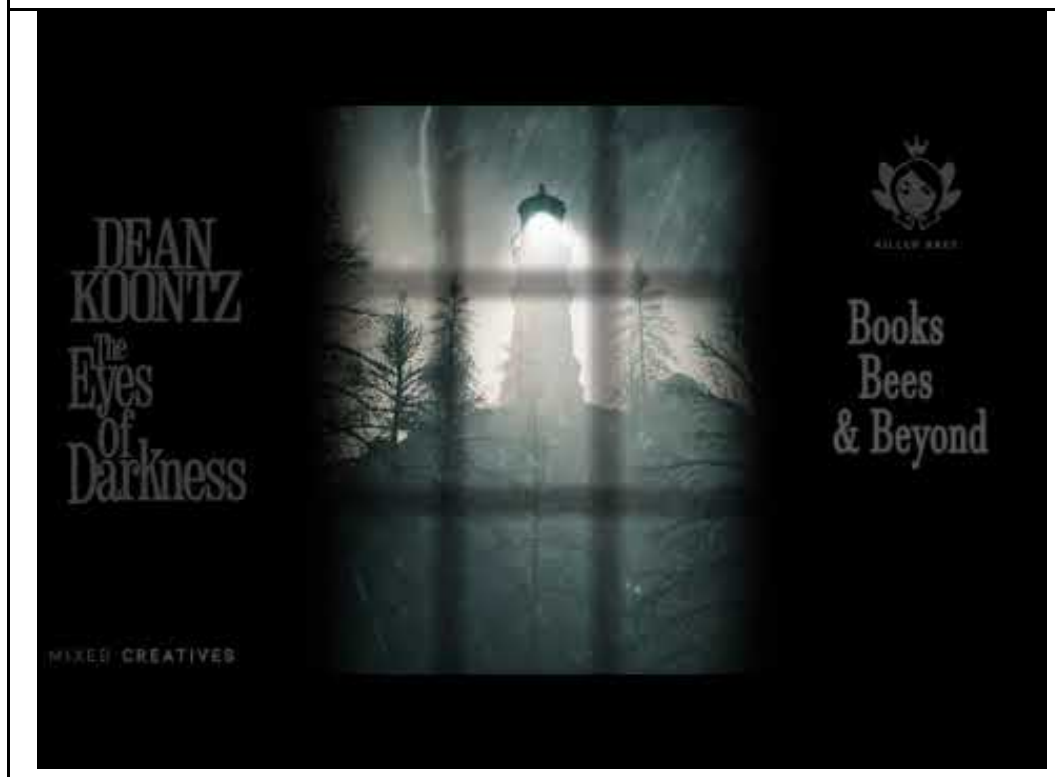

Colorfulness = 2.54 (low extreme)
https://i.ytimg.com/vi/b2OKnU8FVlQ/hqdefault.jpg

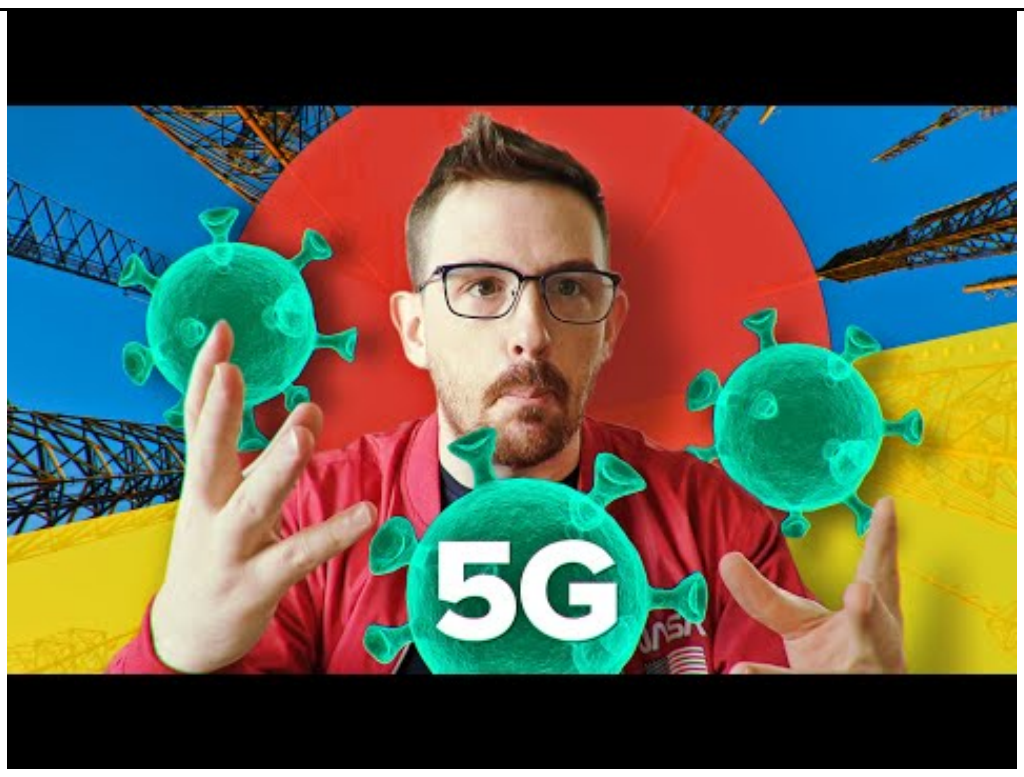

Colorfulness = 42.02 (high extreme)
https://i.ytimg.com/vi/SZqlbWSUi_8/hqdefault.jpg